\documentclass[reprint, superscriptaddress,a4paper,
 amsmath,amssymb, aps,longbibliography]{revtex4-2}

\usepackage{titlesec}
\usepackage[dvipsnames]{xcolor}

\titleformat{\subsection}
  {\normalfont\bfseries}
  {}
  {0em}
  {}
  [] 
  
\titlespacing*{\subsection}{0pt}{1.5ex plus 1ex minus .2ex}{0.5ex plus .2ex}

\usepackage{graphicx}%
\usepackage{multirow}%
\usepackage{amsmath,amssymb,amsfonts}%
\usepackage{amsthm}%
\usepackage{mathrsfs}%
\usepackage[title]{appendix}%
\usepackage{xcolor}%
\usepackage{textcomp}%
\usepackage{upgreek}
\usepackage{siunitx}
\usepackage{verbatim}
\usepackage{hyperref}
\usepackage{csvsimple}
\usepackage{longtable}

\newcommand{\um}[0]{$\upmu$m}
\newcommand{\uJ}[0]{$\upmu$J}
\newcommand{\ka}[0]{K$\upalpha$}

\newcommand{\Te}[0]{$T_\mathrm{e}$}

\newcommand{\kb}[0]{K$\upbeta$}
\newcommand{\kbh}[0]{K$\upbeta_\mathrm{h}$}
\newcommand{\kah}[0]{K$\upalpha_\mathrm{h}$}
\newcommand{\kg}[0]{K$\upgamma$}

\newcommand{\kjcm}[0]{kJ/cm$^2$}
\newcommand{\wcm}[0]{W/cm$^2$}
\newcommand{\figref}{Fig.~\ref}

\newcommand{\hzdr}{Helmholtz Zentrum Dresden Rossendorf, Bautzner Landstra{\ss}e 400, 01328 Dresden, Germany}
\newcommand{\xfel}{European XFEL, Holzkoppel 4, 22869 Schenefeld, Germany}
\newcommand{\prg}{Institute of Physics, Academy of Sciences of the Czech
Republic, Na Slovance 2, 182 21 Prague 8, Czech Republic}
\newcommand{\llnl}{Lawrence Livermore National Laboratory, 7000 East Avenue, Livermore, California 94550, USA}
\newcommand{\desy}{Centre for X-ray and Nano Science CXNS, Deutsches Elektronen-Synchrotron DESY, Notkestrasse 85, 22607 Hamburg, Germany}

\newcommand{\tud}{Technische Universität Dresden, 01062 Dresden, Germany}

\newcommand{\SSL}{Space Science Laboratory, University of California, Berkeley, California 94720, USA}

\newcommand{\eli}{The Extreme Light Infrastructure ERIC, ELI Beamlines Facility, Za Radnici 835,, 25241 Dolni Brezany, Czech Republic}

\newcommand{\ipp}{Institute of Plasma Physics of the CAS, U Slovanky 2525/1a,18200 Prague, Czech Republic}

\begin{document}

\preprint{APS/123-QED}
\title{Plasma screening in mid-charged ions observed by K-shell line emission}

\author{M. Šmíd}
 \email{m.smid@hzdr.de}
\affiliation{\hzdr}
\author{O. Humphries}
\affiliation{\xfel}
\author{C. Baehtz}
\affiliation{\xfel}

\author{V. Bouffetier}
\affiliation{\hzdr}

\author{E. Brambrink}
\affiliation{\xfel}
\author{T. Burian}
\affiliation{\prg}

\author{V. Cerantola}
\affiliation{\xfel}

\author{M. S. Cho}
\affiliation{\llnl}
\author{T.E. Cowan}
\affiliation{\hzdr}
\author{L. Gaus}
\affiliation{\hzdr}
\author{M. F. Gu}
\affiliation{\SSL}
\author{V.~Hájková}
\affiliation{\prg}
\author{L. Juha}
\affiliation{\prg}
\author{J. Kaa}
\affiliation{\xfel}

\author{Z. Konopkova}
\affiliation{\xfel}
\author{M. Kozlová}
\affiliation{\ipp}
\affiliation{\eli}
\author{H. P. Le}
\affiliation{\llnl}
\author{M. Makita}
\affiliation{\xfel}
\author{X. Pan}
\affiliation{\hzdr}
\affiliation{\tud}
\author{T.~Preston}
\affiliation{\xfel}
\author{A. Schropp}
\affiliation{\desy}
\author{J. P. Schwinkendorf}
\affiliation{\hzdr}
\author{H. A. Scott}
\affiliation{\llnl}
\author{R. Štefaníková}
\affiliation{\hzdr}
\author{J. Vorberger}
\affiliation{\hzdr}
\author{W. Wang}
\affiliation{\desy}
\author{U. Zastrau}
\affiliation{\xfel}
\author{K. Falk}
\affiliation{\hzdr}
\affiliation{\prg}

\date{\today}
           
\begin{abstract}
Dense plasma environment affects the electronic structure of ions via variations of the microscopic electrical fields, also known as \emph{plasma screening}. This effect can be either estimated by simplified analytical models, or by computationally expensive and to date unverified numerical calculations. We have experimentally quantified plasma screening from the energy shifts of the bound-bound transitions in matter driven by the x-ray free electron laser (XFEL).  This was enabled by identification of detailed electronic configurations of the observed \ka, \kb\ and \kg\ lines. This work paves the way for improving plasma screening models including connected effects like ionization potential depression and continuum lowering, which will advance the understanding of atomic physics in Warm Dense Matter regime. 

\end{abstract}

\maketitle

Electrons bound in atoms are held at specific levels -- shells and subshells. The energy of these levels is determined by the electric potential of the ion, which is influenced by the presence of other electrons, whether bound within the atom or freely moving in its immediate vicinity. The simplest way to measure this influence is through radiative atomic transitions, i.e., processes in which a bound electron moves from one level to another, accompanied by the emission or absorption of an x-ray photon with a wavelength exactly corresponding to the energy difference of the levels.
Such transitions, including the Cu \ka\ line whose behaviour is studied in this work, were observed and characterized already in 1909 \cite{barkla1909lxix}. Its wavelength (energy) was first measured in 1913 \cite{Moseley} with an deviation better than 3\% to modern benchmarks \cite{xray-data-booklet2001}. Those observations actually led to the discovery of electronic structure of ions. Even today, observing changes in the energies of these transitions remains an excellent method for revealing the structure of atoms and their sensitivity to the surrounding environment. Among other effects, we can speak of line shifts due to two factors: the influence of bound electrons (electron configuration) and the influence of free electrons.

The understanding and quantitative analysis of both shifts relies on complex modelling, which depends on several approximations. One of those is the \emph{plasma screening}, describing how the free electron environment affects the potential of the emitting ion. Most used models of plasma screening are based on the calculations from about 60~years ago \cite{EK,SP}. The advent of x-ray free electron lasers (XFELs) opened up new possibilities to experimentally challenge those models, to observe plasma screening via a shift of the emission lines or absorption edges as a function of plasma conditions. Still, it is typically not straightforward to extract the \emph{continuous} effect of screening, since the line shifts are at the same time influenced by the \emph{discrete} changes of the electronic configuration, which is, similarly as screening, affected by plasma temperature.  

The change of line position due to bound electrons, or, in other words, electronic configuration, can be well illustrated on the $1s-2p$ transition. Its energy is mostly influenced by the K-shell occupation, electrons in the L-shell have a smaller effect, and the influence of M- and N- shell electrons can be often neglected: The addition of a K-shell electron actually changes the name of the transition: so $1s^1-2p^1$ is called Ly$_\upalpha$, while additional electron leads to $1s^2 - 1s2p$ which is the He$_\alpha$. Adding an electron to the L-shell can produce $1s^2 2p - 1s2p^2$, which could be called either Li-like satellite of He$_\alpha$, or Li-like \ka. Transitions from ions with more electrons can then be called \ka\ satellites. Those were first calculated in 1969 \cite{House_1969} as a function of the L-shell occupation, and experimentally observed in 1975 \cite{Jamison_1975}. However, the exact description of many-electron systems like Cu is computationally challenging, as the number of possible configurations of available electrons (29 in copper) is vast, and the lines are mostly indistinguishable. Such modelling of non-LTE plasmas was attempted with super-configuration codes, but the results still show large deviations from observations \cite{Lee2023}. For example, in the atomic model presented in this paper, the number of K shell transitions is about 120 million, out of which about 1.5 million represents $1s-2p$ lines. In experiment, typically only 9 emission lines defined by the occupation of the L-shell are resolvable. These \emph{heated} \ka\ satellite lines have vast applications at Warm Dense Matter (WDM) and plasma diagnostics, as identified already in 1981 \cite{Nardi}, their recent applications are shown e.g. in \cite{Chen-2009-xray, Smid2019, Ordyna, kluge2016, Deschaud2020}. The advantage of those lines is that their emission is produced by highly charged ions present in temperatures of hundreds to thousands of eV, while their emission energies lie in a narrow range well resolvable by high-resolution crystal spectrometers, therefore providing a unique insight into the plasma conditions.  The proper understanding of x-ray spectroscopy and atomic physics is also important for fusion research \cite{seely2001}.

The second effect altering the line position is the \emph{plasma screening} \cite{Hu2022}, occurring when the free electrons surrounding the ion alter its electric field. The screening has several consequences: 
The change of transition energy is often called \emph{Stark shift}. The \emph{ionization potential depression} (IPD) describes the decrease of energy needed to remove an bound electron into continuum, most often manifested in the form of shift of the absorption edge. \emph{Continuum lowering} (CL) \cite{more1982} shows that the boundary between free electrons (continuum) and bound ones is decreasing, and therefore outer shells are effectively disappearing -- merging into continuum. These effects are most often described by the Stewart-Pyatt model (SP) \cite{SP} from 1966. Some of recent experiments identifying the shifts of the K-edge indicated that the modified Ecker-Kr\"oll model \cite{EK} fit the measurements better \cite{Hu2022,Ciricosta2012,Ciricosta2016} or where the SP model fits closer \cite{Hoarty}, spurring further model development \cite{Hu2022, Benredjem, Zan2021} and discussions \cite{Iglesias2014}.

Pioneering experiment studying continuum lowering in atoms in dense plasma environment isochorically heated by XFEL beam at the LCLS laboratory have been shown in 2012 \cite{Ciricosta2012}. The shift of absorption edges was measured in low-Z materials (Al, Mg, Si). The shift in Al was later calculated by using Density functional theory (DFT), with results in perfect agreement to the experimental data with charge states 3--7 \cite{vinko2014density}.
One recent approach to quantify IPD was shown in \cite{Benredjem}, where the electron distribution was modelled by classical molecular dynamics. The averaged effect over an ensemble of configurations was then calculated, and shown to agree well with the previous experimental data up till charge state 9. However, the spectral simulations can completely skip the concept of IPD, as shown in \cite{Hu2022}. Here, the DFT-based multi-band kinetic model (VERITAS) explicitly accounts for the interactions among ions and the dense plasma environment. Energy band shifting and ionization balance are therefore self-consistently calculated, without invocation of an ad hoc CL or IPD model. Such models are extremely computationally expensive and therefore it might be difficult to apply them to the complex atomic structures like those presented in this paper.

In 2012, Hu \textit{et al.} pointed out that \emph{``Detailed spectroscopic measurements at warm dense matter conditions are rare, and traditional collisional-radiative equilibrium models, based on isolated-atom calculations and ad hoc continuum lowering models, have proved questionable at and beyond solid density."} \cite{Hu2022} In this paper, we provide the experimental data and extract the measurement of plasma screening in high energy density regime. 
We show x-ray emission spectra from copper driven by narrow bandwidth XFEL pulses, with sufficient intensity to heat and ionize the material, generating an array of transitions within the pulse length, including double core hole states. The tunability of the XFEL photon energy is used to resonantly pump transitions with known electronic configuration. By careful analysis of those resonances and comparison to a detailed model of K-shell transitions calculated by the Flexible Atomic Code \cite{Gu2008}, we connect measured emission lines with the charge state and L-shell occupancy, and consequently measure their shift compared to calculated values. The same transitions are also calculated while applying plasma screening by the SP model \cite{Gu2020} to show its difference to the experiment. The unique feature of this dataset is that the measurement contains the \ka, \kb, and \kg\  transitions, therefore describing the modification of all electronic shells present in the material. The observed shifts are shown to \emph{not} agree well with the Stewart-Pyatt model. The data show the complex structure of the K-shell emission in highly charged ions, and aim to guide the future development and verification of new, more precise, models.

 \begin{figure}[bht]
\begin{center}
\includegraphics[width=\linewidth]{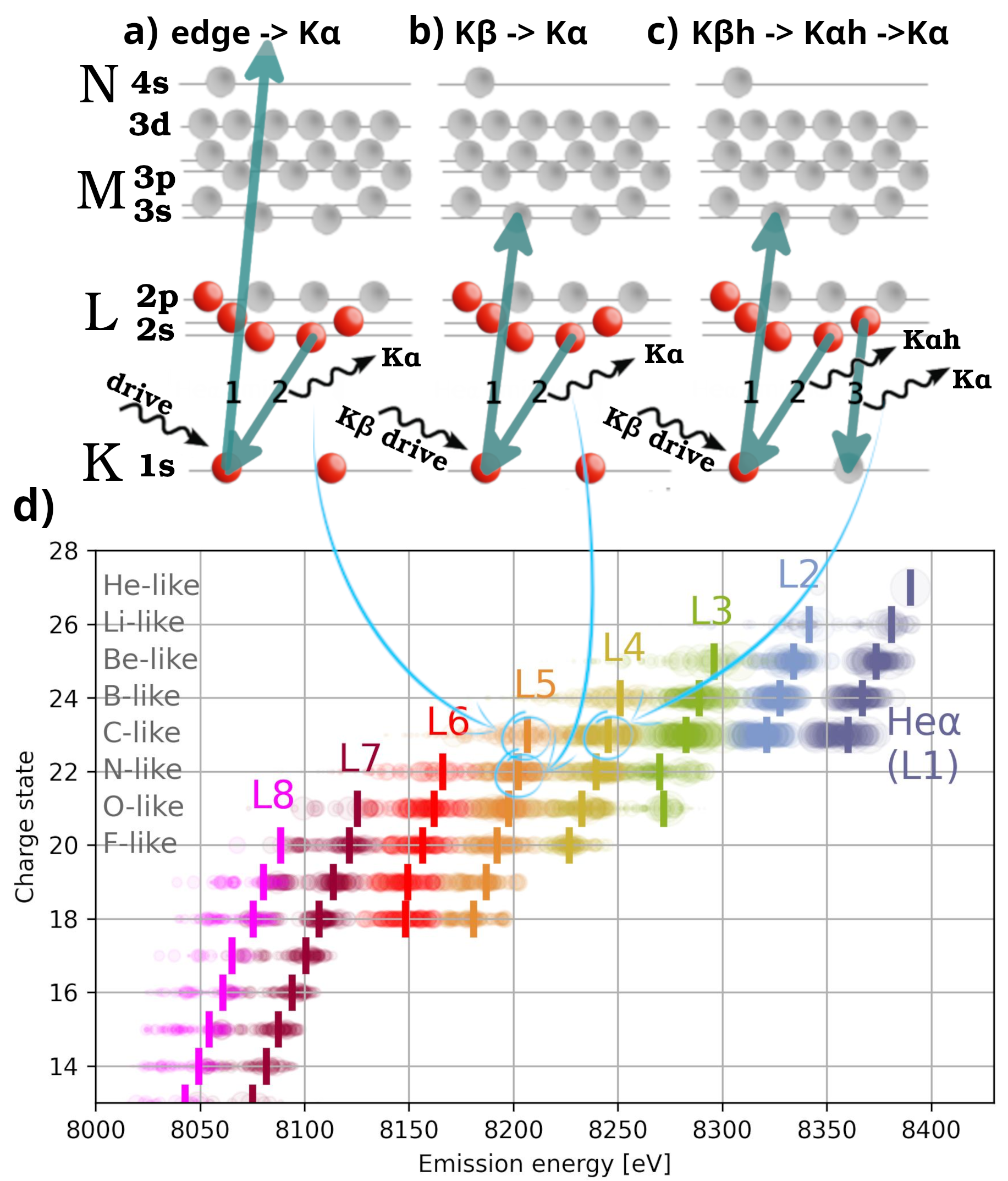}
\caption{\label{kadetail}
Schematic depiction of observed transition chains: emission above K edge (a), \ka\ emission driven by \kb\ absorption (b), and \ka\ and \kah\ driven by \kbh\ (c).
\ka ~transitions calculated by the FAC code (d). Each circle is a single transition with size corresponding to its oscillator strength; the transitions are grouped according to charge state (y-axis) and L-shell occupancy (color), and a weighted mean for each group is shown by a vertical marker. 
}
\end{center}
\end{figure}

\subsection{K-shell transitions in calculations}

The K-shell energies and oscillator strengths of the emission lines were calculated by the FAC code \cite{Gu2008}, details of the calculation are in Appendix. The $1s^2-2p^n$ transitions are shown in \figref{kadetail} (d).  A mean energy weighted by the oscillator strength is calculated for each group of transitions given by charge state and L-shell occupancy.

The dominant factor affecting the energy is the occupancy of the L-shell, therefore we label the groups as  \ka~L$x$, where $x$ is the L-shell occupancy in the upper state. 
Such a description, however, is insufficient to comprehend the full dynamics, as shifts of each of those lines as a function of charge state (or occupancy of M-shell) are resolved. Therefore using a nomenclature  \ka\ L$x$\ M$y$ might be necessary in specific cases.  
The lines with various M shell occupancies within given \ka\ L$x$ transition are typically unresolvable in experimental data, as their shift is smaller than their widths. In the data shown in this paper, the distinction was made possible by selectively pumping various excitation and ionization states via the \kb\ transitions, whose sensitivity to M shell occupancy is significantly higher.

Similar data as in \figref{kadetail} (d) are also calculated for other transitions of interest, namely the \ka\ transition in so-called hollow ions (ion with a hole in K-shell in the initial state) \cite{rosmej2007}, further as \kah~ ($1s^1-2p^\mathrm{N} $), \kb\ ($1s^2-3p^\mathrm{N} $), \kbh~ ($1s^1-3p^\mathrm{N} $), and \kg~($1s^2-4p^\mathrm{N} $), corresponding figures are shown in Extended data. 
 Further FAC calculations were run with the plasma potential modelled by the SP model, assuming solid density and various plasma temperatures. 

In order to demonstrate the basic scaling of those line shifts, an empirical formula is designed to approximate the simulated line positions. 
This says that the energy of transition can be approximated as
\[E = E_0 - k_\mathrm{K}K- k_\mathrm{L}L - k_\mathrm{M}M - c/Te,\] 

where $K$, $L$, and $M$ are the occupancies of respective shells, \Te~is the plasma electron temperature in eV, and $E_0$, $k_\mathrm{K}$, $k_\mathrm{L}$,$k_\mathrm{M}$, and $c$ are constants summarized in Tab.\ref{consts}.
The $k_\mathrm{x}$ constants indicate how much the emission line shifts with addition of one electron into given shell. Addition of an electron into the L shell introduces a shift of about 44 and 98 eV for \ka ~and \kb, respectively, while the M shell electron causes shifts of only 4, respectively 15 eV. Those shifts are slightly increased for ions with close-to-full L shells. The sensitivity to temperature is significantly higher for \kb\ compared to the \ka\ transition. The fit is valid only for charge state discussed in this work, i.e. between 13 and 27. 

\begin{table}
    \centering  
    \begin{tabular}{lccccc}
    \hline\hline 
   Case & $E_0$ & $k_\mathrm{K}$  & $k_\mathrm{L}$ & $k_\mathrm{M}$ & $c$ \\
    \hline
        \ka ~($L\le6$)&~~ 9025~~& 300& ~~44~~ & ~~4~~ &~~ 200 ~~\\
        \ka ~($L\ge7$)&~~9025 & 300 & 43 & 6 & 100 \\
        \kb ~($L\le6$) & ~~10630 & 380 & 98 & 15 & 500 \\
        \kb ~($L\ge7$)& ~~10630 & 380 & 96 & 19 & 300 \\
        \kg 	                 & ~~11110 & -- & 118 & 30 & --- \\
    \hline
    \end{tabular}
    \caption{Fit parameters for simple formula of \ka~ and \kb~ energies.}
    \label{consts}
\end{table}

A similar dependence of the transition energy on the presence of electrons in the M-shell (often called spectator electrons) was shown e.g. for Mg He$_\alpha$ in \cite{perezcallejo-2023-dielectronic}. To our knowledge, however, such shifts were not quantitatively resolved and described before for ions with more than 3 electrons.

\begin{figure*} [hbt]
\begin{center}
\includegraphics[width=\linewidth]{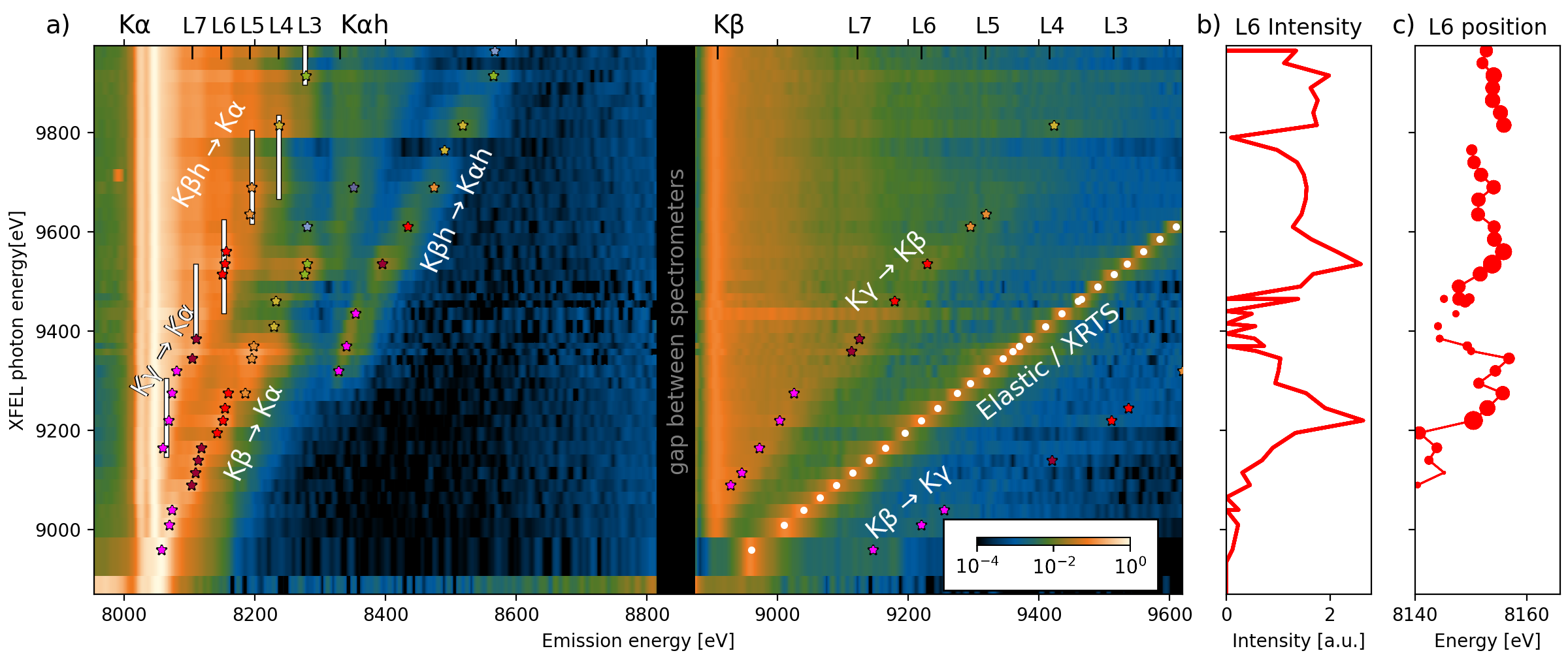}

\caption{\label{abeled} Experimental spectra for beam energy density 110 \kjcm (a) with identified resonances (stars with color corresponding to L-shell occupancy), edges (white bars), and elastic scattering (white circles). The fitted intensity and position of \ka\ L6 is shown in (b) and (c).}
\end{center}
\end{figure*}

\subsection{Experiment}
The experiment was performed at the HED instrument of the European XFEL laboratory \cite{ZastrauHED}. The 25 fs long x-ray beam was focused down to a sub-\um ~focal spot reaching intensities up to \num{7e18} \wcm, corresponding to irradiation of \num{180} \kjcm, and its photon energy was varied in the wide range above Cu K edge (8.9 -- 9.9~keV). X-ray emission of the 3~\um\ thick Cu foil was measured by crystal spectrometers covering the range from neutral \ka\ till highly charged \kb\ transitions. Spectra were measured for a variable incident pulse energy, and the knowledge of the focal spot distribution allowed to perform the \emph{focal spot inversion} (see Methods) to extract the spectra for a given irradiation (areal energy density). The spectra extracted for irradiation of 110 \kjcm\ and various XFEL photon energies are shown in \figref{abeled} (a). The markers are showing the three key features, which are subject of further analysis: X-ray Thomson scattering (XRTS), resonances, and edges. 

\subsection{Resonances}
The resonant processes mean a chain of two or more electronic transitions, where the first one is driven by photoexcitation, ended by a radiative de-excitation. As the de-excitation follows typically on a few fs timescale (as observed in our simulations), the probability of another process modifying the configuration in between is low in the present cases, therefore we assume the state of the ion is otherwise unchanged. Two such processes are depicted in \figref{kadetail} (b, c).

The resonant processes are identified as emission peaks in the spectra for a particular driving energy, whose intensity is decreasing if the driving energy is changed. Measures of the emission were constructed via fitting a family of Gaussian peaks to the measured spectrum, giving an estimate of yield and position. The details of the fit and example of the spectral lineouts are shown in Appendix \ref{fitting}.
By applying the central limit theorem to the combinatorial nature of spectator electrons, these peaks should be well represented by a Gaussian given the approximately linear dependence of the spectator electron perturbations.
The intensity and position of the fits of \ka ~L6 line are shown in \figref{abeled} (b) and (c).
The two maxima in the intensity plot show the resonant driving via \kb~ (at $\approx~9250$ eV) and \kbh~ (at $\approx 9500$ eV).  In both cases, the emission energy shifts with the change of the driving energy, because different charge states of \kb, respectively \kbh, are being pumped, therefore producing \ka\ emission of ions with corresponding charge states.  Once the driving energy moves above $\approx 9600$ eV, both the intensity and position of the line is not changing significantly, because it is emitted from ions with K-hole made by photoionization - \figref{kadetail} (a) - not by photoexcitation. 

\begin{figure*}[phtb]
\begin{center}
\includegraphics[width=\linewidth]{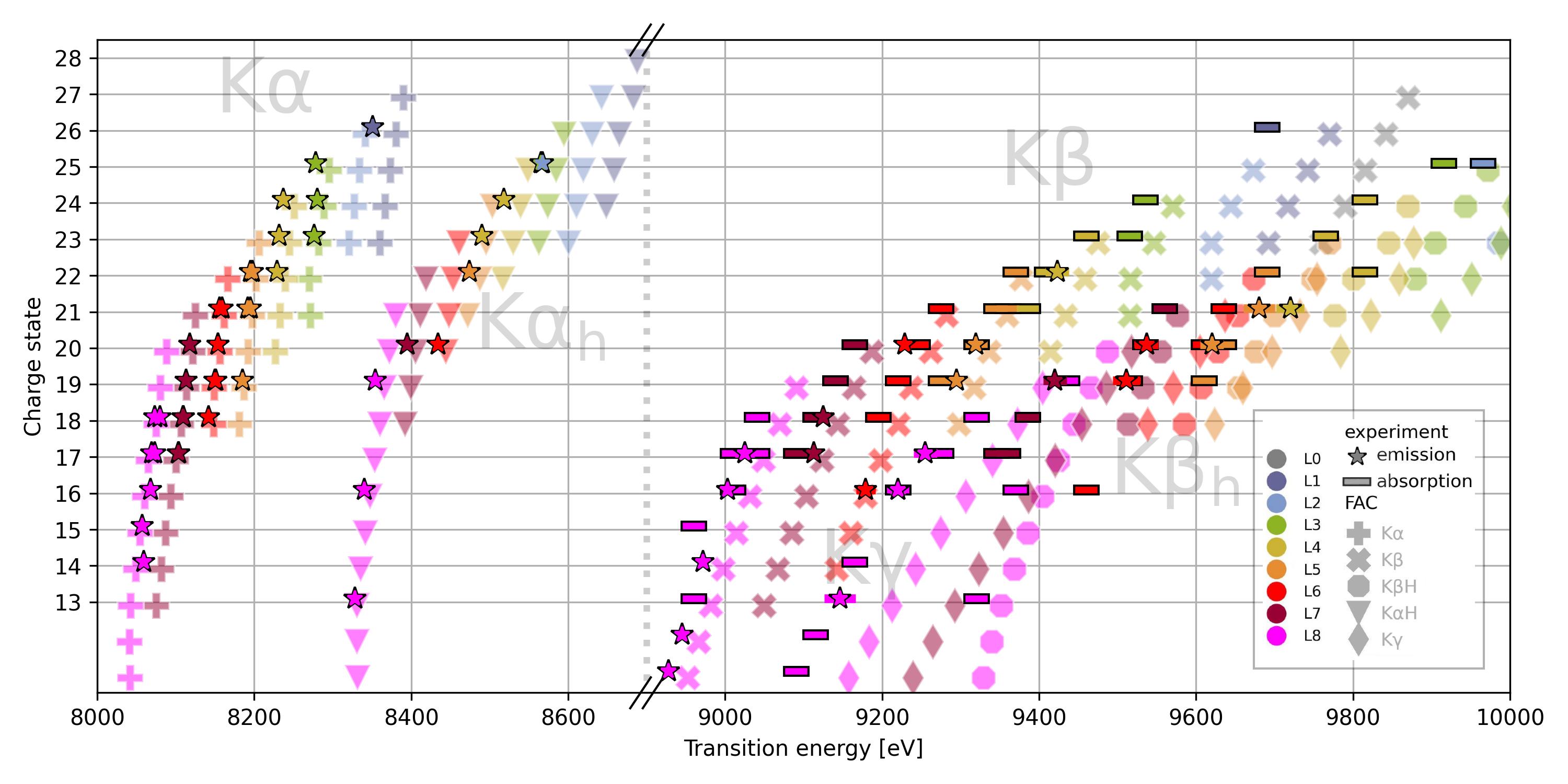}
\caption{\label{map} Map of spectral lines in Cu. Black outlined stars show emission lines and bars absorption energies observed in the experiment. 
 Translucent symbols are calculated by the FAC code for isolated atom.
Color of symbols corresponds to L shell occupancy.}
\end{center}
\end{figure*}

To identify the electronic configurations of the observed transitions, their energies were compared to the theoretical model, as shown in \figref{map}. Each experimentally observed point from \figref{abeled} is interpreted as a pair of 'driving line --- emission line', where the driving line is represented by the horizontal bar (with width of 25 eV showing the uncertainty given by XFEL bandwidth) and the emission by the black-outlined circles. To each measured pair of emission --- absorption channels, the charge state and L-shell occupancy is assigned by identifying the theoretical pair with best matching energies.


\subsection{Edges}
Identification of the absorption edges has a long tradition in this type of data \cite{Ciricosta2012}, and is typically the easiest measurement that could be done. In copper as well as in other materials with similar Z, however, the edge position is sharing similar energies as the \kbh~line, undermining the capability to estimate the edge position precisely. Still, we have experimentally identified a relatively broad range of driving photon energies within which the absorption edge can lie, those ranges are shown as white vertical lines in \figref{abeled} (a). The bottom edge of the range is identified so, that if the driving photon energy go below this edge, the emission of the corresponding \ka\ line significantly weakens compare to values above it. 
The upper boundary is the lowest energy above which the emission does not show significant intensity or energy fluctuations. As mentioned, the ranges overlap with the \kbh\ -- \ka~resonances (\figref{abeled}(a)), which is the main reason why this data does not allow a more precise estimation of the edge position. The rebinding of states from the continuum, which retain broad energy bands due to the extent of their wavefunction, has been found to make clear identification of an edge position challenging, resulting in different conclusions on the required IPD \cite{Ciricosta2016,Gawne2023,perezcallejo-2023-dielectronic}.

\begin{figure*}[phtb]
    \centering
    \includegraphics[width=1\linewidth]{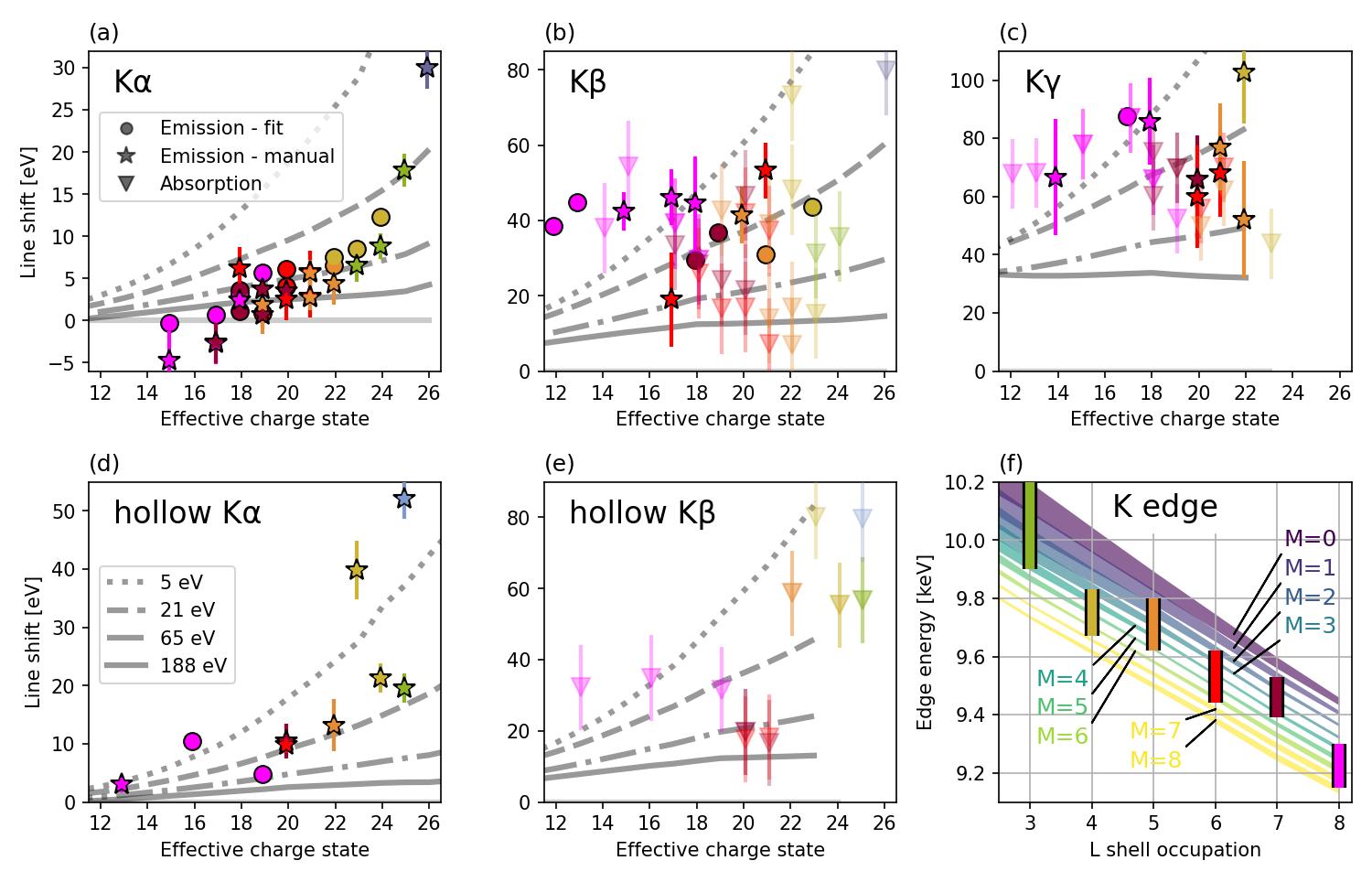}
    \caption{Observed line shifts (a...e) and edge positions (f) for energy density 110 \kjcm. Stars are measured by emission, triangles by absorption with errorbar corresponding to the XFEL bandwidth. The color corresponds to L shell occupancy with same coding as Fig. 3. Grey lines are predictions by SP model in FAC for various temperature assumptions, labeled in (d) for all panes. The bands in (f) show the calulated edges for various M-shell occuapncy, width of band contains data for temperaterus between 5 and 107 eV. } 
    
    
    \label{screening}
\end{figure*}

Since the M-shell occupancy and therefore the charge state in the measurement of the K edge is unknown, the resulting data in \figref{screening} are shown as a function of L shell occupancy. The FAC model of the edge position is then plotted  by bands for various M-shell occupancies with different colors. Those bands are broad to contain edges for the temperatures between 5 and 107 eV. The effect of temperature is shown to be smaller then effect of M shell electrons. The experimental data agree to models with 1 \dots 7 electrons in M shell. The identification of resonant transitions is showing typically between 0 \dots 4 M shell electrons, therefore the edge measurement also indicates the screening and CL is stronger then predicted by SP model.

\subsection{Charge-state dependence of plasma screening}
In this work, plasma screening is measured as a difference between the observed transition energy and calculation of its energy for isolated atom.
The found values are plotted in \figref{screening} for each transition separately as a function of effective charge state. Effective charge state in this context is charge state calculated from occupancy of K, L and M shells, i.e. ignoring electrons in N shell, which have negligible effect on screening, but in studied case appeared due to \kg\ excitation. Plasma screening calculated by the FAC code using the Stewart-Pyatt model is shown in grey lines, assuming solid density and different plasma temperatures.  

The trend that plasma screening is increasing with charge state agrees. 
However, it is not following the trend lines and the models would have to assume lower temperatures (between 5 and 20~eV), which is significantly lower than the expected values of more than a hundred eV, similar to other x-ray isochoric heating investigations for similar states \cite{perezcallejo-2023-dielectronic}.  

The K edge is shown in \figref{screening}(f). Each band shows variation of edge position for the temperatures 5\dots 188 eV, showing that the temperature effect is negligible. On the other hand, the effect of M-shell occupancy is strong, each electron in M band decreasing the edge by 50 eV. We can therefore conclude that the experiment agrees to model only if M-shell occupation is between 2 to 6 electrons for L=4\dots7, and more then 5 electrons for L=8. Those numbers are not that unreasonable, but from the identification of resonances, we expect less electrons in M shell, therefore indicating that the modelled shift is underestimated.


\subsection{Thermal conditions}

\begin{figure}
    \centering
    \includegraphics[width=1\columnwidth]{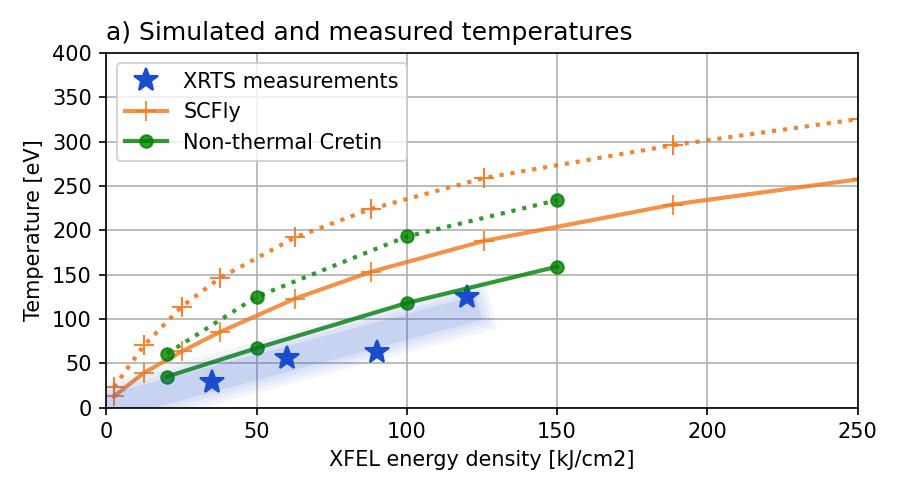}
    
    \includegraphics[width=1\columnwidth]{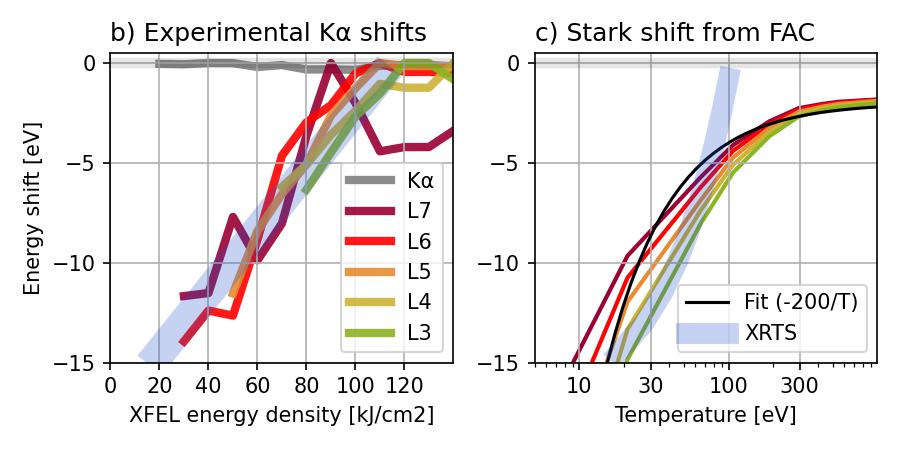}

    \caption{Temperatures of the plasma calculated by the CR codes SCFLY  and Cretin (a). Solid lines are temperatures during the peak of the XFEL beam, dotted lines are the maximal temperatures, reached toward the end of the pulse. Stars indicate temperatures estimated from the XRTS data. Observed (b) and predicted (c) Stark shifts as a function of energy density and plasma temperature, respectively. Blue band in (b) is a linear fit to the data, and is transferred into (c) by using the measured temperature-energy density relation in (a).}
    \label{temperatures}
\end{figure}

There are several ways  to assess the thermal conditions in the target. First, a set of simulations with the SCFLY collisional radiative (CR) code \cite{Chung2017} was performed. The inherent disadvantage of that code is that it assumes thermal distribution of electrons, which in general might not be true, as the dominant mechanism of energy absorption is photoionization, producing electrons with very non-thermal energies. However, as shown in \cite{Ren2023nat_comm} and confirmed by our simulations with the non-thermal version of the code Cretin, the electron distribution in this case is close to Maxwellian due to rapid thermalization via frequent electron--electron collisions, allowing the temperature to be used as a suitable metric. The Cretin code was run with comparable conditions, and the temperatures obtained from both codes are presented in \figref{temperatures} (a). 

The experimental approach analyses the elastic scattering data (XRTS) using the approach shown in \cite{Dornheim2022}. The temporal integration of the signal, overlap of XRTS with emission, and overall signal to noise ratio, however, limits its accuracy. The results are shown in \figref{temperatures} (a) and are in good agreement to the simulated values for temperatures during the peak irradiation, concluding that the presented data with energy density 110 \kjcm 
were measured in plasma with $T \approx 100 \dots 150$~eV.

\subsection{Temperature dependence of plasma screening}
All experimental spectra shown until this point were obtained with an irradiation of 110 \kjcm. Investigations of line positions from different heating conditions can reveal the Stark shifts as a function of temperature. The shifts of the \ka~Lx emission extracted from spectra with various XFEL energy densities is shown in \figref{temperatures} (b). All transitions show very similar trend -- about 13 eV shift between 40 and 120 \kjcm. Such shift corresponds to the theoretical model of plasma screening (\figref{temperatures}c) with temperature of about 20 eV. Note, that in contrast to common intuition, the higher temperatures correspond to more equilibrated conditions in this situation: We observe lines with charge state 22 or more, which would be present in equilibrated plasmas only at much higher temperatures (one or few keV);
The lower the temperature we observe those transitions at, the further the conditions are from any kind of equilibria, and the stronger the Stark shift is. 

Having a reliable Stark shift model for those conditions, those shifts could be used to infer the electron temperature of the plasma. 
Yet, an inverse process can be applied here: Observed line shifts are ascribed to plasma temperature from the XRTS measurements, and therefore an empirical curve of Stark shift as a function of temperature is plotted in \figref{temperatures}(c) with broad blue line. This agrees reasonably well with the theoretical prediction.

\subsection{Summary}
We have mapped the K shell transitions in intermediate to highly charged matter by using resonant pumping by the high intensity XFEL beam. The connection between the absorbing and emitting line allowed identification of the L and M shell occupancy, and therefore resolution of the transitions with unprecedented details, i.e. to distinguish its emission for various occupancies of K, L \emph{and} M shells. The FAC atomic calculations of those transitions in isolated ions, as well as with plasma screening by the Stewart-Pyatt model with various plasma temperature assumptions were compared to the experimental observation to identify the discrepancies. 
In both the charge-state and temperature dependent measurement, experimental data are found to be matched only by the SP model when assuming unreasonably low temperatures - reconfirming the observation that this model systematically underestimates the plasma screening in this regime.
This statement was so far observed only for charge states below 15, our data confirms it for charge state up to 26. The presented experimental measurements of shifts of \ka, \kb, \kg\ transitions and their hollow partners 
provides a complex information about the modification of the ionic electronic potential in a well defined plasma environment. These measurements shall stimulate the development and verification of novel codes to model the potential and improve our understanding of precise atomic physics in Warm Dense Matter.

\subsection{Data Availability}
The raw data are published by the XFEL laboratory, see \cite{xfel2806}, the processed data \cite{rodare_data} and the analysis scripts \cite{rodare_scripts} are published in Rodare repository.

\subsection{Acknowledgments}
We acknowledge European XFEL in Schenefeld, Germany, for provision of X-ray free-electron laser beamtime at Scientific Instrument HED  (High Energy Density Science) and would like to thank the staff for their assistance. The work was also supported by the Helmholtz Association under the grant no. VH-NG-1338. The work of M.S.C., H.P.L. and H.A.S. was performed under the auspices of the U.S. Department of Energy by Lawrence Livermore National Laboratory under Contract DE-AC52-07NA27344.

\subsection{Author contributions}
M.Š. wrote the manuscript, performed majority of the data analysis and run the FAC and SCFly simulations. O.H. aided the data analysis and writing, R.Š., X.P., H.P.L., A.S., J.V and T.E.C. edited and reviewed the manuscript. The experiment was lead by M.Š., K.F., O.H., M.K. and C.B., and performed by E.B., V.B., V.C., L.G., Z.K., M.M. X.P., T.P., J.P.S., R.Š. and U.Z.; T.B., V.H, L.J. performed and analysed the focal spot via the imprint technique. A.S. and W.W. helped with the x-ray focusing. The support with atomic simulations was provided by M.S.Ch. and H.A.S.; M.F.G advised on FAC and data analysis, H.P.L. provided the cretin simulations. J.V. did the XRTS analysis.


\bibliographystyle{unsrt}
\bibliography{bibliography}

\begin{thebibliography}{10}

\bibitem{barkla1909lxix}
Charles~Glover Barkla and Ch~A Sadler.
\newblock Lxix. the absorption of r{\"o}ntgen rays.
\newblock {\em The London, Edinburgh, and Dublin Philosophical Magazine and
  Journal of Science}, 17(101):739--760, 1909.

\bibitem{Moseley}
H.G.J. Moseley.
\newblock Xciii. the high-frequency spectra of the elements.
\newblock {\em The London, Edinburgh, and Dublin Philosophical Magazine and
  Journal of Science}, 26(156):1024--1034, 1913.

\bibitem{xray-data-booklet2001}
Albert~C. Thompson, Douglas Vaughan, David~T. Attwood, Eric~M. Gullikson,
  Malcolm~R. Howells, Jeffrey~B. Kortright, Arthur~L. Robinson, James~H.
  Underwood, Kwang-Je Kim, Janos Kirz, Ingolf Lindau, Piero Pianetta, Herman
  Winick, Gwyn~P. Williams, and James~H. Scofield.
\newblock {\em X-Ray Data Booklet}.
\newblock Center for X-Ray Optics, Lawrence Berkeley National Laboratory, 2nd
  edition, 2001.
\newblock Accessed: 2025-10-02.

\bibitem{EK}
G.~Ecker and W.~Kröll.
\newblock {Lowering of the Ionization Energy for a Plasma in Thermodynamic
  Equilibrium}.
\newblock {\em The Physics of Fluids}, 6(1):62--69, 01 1963.

\bibitem{SP}
John~C. {Stewart} and Jr. {Pyatt}, Kedar~D.
\newblock {Lowering of Ionization Potentials in Plasmas}.
\newblock {\em \apj}, 144:1203, June 1966.

\bibitem{House_1969}
Lewis~L. House.
\newblock {Theoretical Wavelengths for Ka-TYPE X-Ray Lines in the Spectra of
  Ionized Atoms (carbon to Copper)}.
\newblock {\em The Astrophysical Journal Supplement Series}, 18:21, February
  1969.

\bibitem{Jamison_1975}
K.~A. Jamison, C.~W. Woods, Robert~L. Kauffman, and Patrick Richard.
\newblock $k\ensuremath{\alpha}$ satellite x rays in al, sc, and ti following
  bromine-ion bombardment.
\newblock {\em Phys. Rev. A}, 11:505--508, Feb 1975.

\bibitem{Lee2023}
Hae~Ja Lee, Sam Vinko, Oliver Humphries, Eric Galtier, Ryan Royle, Muhammad
  Kasim, Shenyuan Ren, Roberto Alonso-Mori, Phillip Heimann, Mengning Liang,
  Matt Seaberg, Sébastien Boutet, Andrew~A Aquila, Shaughnessy Brown, Mikako
  Makita, and European XFELs.
\newblock Driving iron plasmas to stellar core conditions using extreme x-ray
  radiation.
\newblock {\em https://doi.org/10.21203/rs.3.rs-3129538/v1}, 2023.

\bibitem{Nardi}
E.~Nardi and Z.~Zinamon.
\newblock {Ka satellite spectra as diagnostics for particle beam‐target
  interaction}.
\newblock {\em Journal of Applied Physics}, 52(12):7075--7079, 12 1981.

\bibitem{Chen-2009-xray}
S.~N. Chen, P.~K. Patel, H.-K. Chung, A.~J. Kemp, S.~Le~Pape, B.~R. Maddox,
  S.~C. Wilks, R.~B. Stephens, and F.~N. Beg.
\newblock X-ray spectroscopy of buried layer foils irradiated at laser
  intensities in excess of 10 20 w/cm2.
\newblock {\em Physics of Plasmas}, 16(6):062701, 06 2009.

\bibitem{Smid2019}
M.~{\v{S}}m{\'i}d, O.~Renner, A.~Colaitis, V.~T. Tikhonchuk, T.~Schlegel, and
  F.~B. Rosmej.
\newblock Characterization of suprathermal electrons inside a laser accelerated
  plasma via highly-resolved ka-emission.
\newblock {\em Nature Communications}, 10(1):4212, Sep 2019.

\bibitem{Ordyna}
P.~et~al. Ordyna.
\newblock isualizing plasmons and ultrafast kinetic instabilities in
  laser-driven solids using x-ray scattering.
\newblock 2024.

\bibitem{kluge2016}
T.~Kluge, M.~Bussmann, H.-K. Chung, C.~Gutt, L.~G. Huang, M.~Zacharias,
  U.~Schramm, and T.~E. Cowan.
\newblock {Nanoscale femtosecond imaging of transient hot solid density plasmas
  with elemental and charge state sensitivity using resonant coherent
  diffraction}.
\newblock {\em Physics of Plasmas}, 23(3):033103, 03 2016.

\bibitem{Deschaud2020}
B.~Deschaud, O.~Peyrusse, and F.~B. Rosmej.
\newblock Simulation of xfel induced fluorescence spectra of hollow ions and
  studies of dense plasma effects.
\newblock {\em Physics of Plasmas}, 27, 6 2020.

\bibitem{seely2001}
John Seely, Christina Back, Richard Deslattes, Lawrence Hudson, Glenn Holland,
  Perry Bell, and Michael Miller.
\newblock Hard x-ray spectrometers for the national ignition facility.
\newblock {\em Review of Scientific Instruments}, 72(6):2562--2565, 2001.

\bibitem{Hu2022}
Suxing Hu, David Bishel, David Chin, Philip Nilson, Valentin Karasiev, Igor
  Golovkin, Ming Gu, Stephanie Hansen, Deyan Mihaylov, Nathaniel Shaffer, Shuai
  Zhang, and Timothy Walton.
\newblock Probing atomic physics at ultrahigh pressure using laser-driven
  implosions.
\newblock {\em Nature Communications}, 13, 11 2022.

\bibitem{more1982}
Richard~M. More.
\newblock Electronic energy-levels in dense plasmas.
\newblock {\em Journal of Quantitative Spectroscopy and Radiative Transfer},
  27(3):345--357, 1982.

\bibitem{Ciricosta2012}
O.~Ciricosta, S.~M. Vinko, H.~K. Chung, B.~I. Cho, C.~R.D. Brown, T.~Burian,
  J.~Chalupský, K.~Engelhorn, R.~W. Falcone, C.~Graves, V.~Hájková,
  A.~Higginbotham, L.~Juha, J.~Krzywinski, H.~J. Lee, M.~Messerschmidt, C.~D.
  Murphy, Y.~Ping, D.~S. Rackstraw, A.~Scherz, W.~Schlotter, S.~Toleikis, J.~J.
  Turner, L.~Vysin, T.~Wang, B.~Wu, U.~Zastrau, D.~Zhu, R.~W. Lee, P.~Heimann,
  B.~Nagler, and J.~S. Wark.
\newblock Direct measurements of the ionization potential depression in a dense
  plasma.
\newblock {\em Physical Review Letters}, 109, 8 2012.

\bibitem{Ciricosta2016}
O.~Ciricosta, S.~M. Vinko, B.~Barbrel, D.~S. Rackstraw, T.~R. Preston,
  T.~Burian, J.~Chalupský, B.~I. Cho, H.~K. Chung, G.~L. Dakovski,
  K.~Engelhorn, V.~Hájková, P.~Heimann, M.~Holmes, L.~Juha, J.~Krzywinski,
  R.~W. Lee, S.~Toleikis, J.~J. Turner, U.~Zastrau, and J.~S. Wark.
\newblock Measurements of continuum lowering in solid-density plasmas created
  from elements and compounds.
\newblock {\em Nature Communications}, 7, 5 2016.

\bibitem{Hoarty}
D.J. Hoarty, P.~Allan, S.F. James, C.R.D. Brown, L.M.R. Hobbs, M.P. Hill,
  J.W.O. Harris, J.~Morton, M.G. Brookes, \~R. Shepherd, J.~Dunn, H.~Chen,
  E.~{Von Marley}, P.~Beiersdorfer, H.K. Chung, R.W. Lee, G.~Brown, and
  J.~Emig.
\newblock The first data from the orion laser; measurements of the spectrum of
  hot, dense aluminium.
\newblock {\em High Energy Density Physics}, 9(4):661--671, 2013.

\bibitem{Benredjem}
Djamel Benredjem, Jean-Christophe Pain, Annette Calisti, and Sandrine Ferri.
\newblock Plasma density effects on electron impact ionization.
\newblock {\em Phys. Rev. E}, 108:035207, Sep 2023.

\bibitem{Zan2021}
Xiaolei Zan, Chengliang Lin, Yong Hou, and Jianmin Yuan.
\newblock Local field correction to ionization potential depression of ions in
  warm or hot dense matter.
\newblock {\em Physical Review E}, 104, 8 2021.

\bibitem{Iglesias2014}
Carlos~A. Iglesias.
\newblock A plea for a reexamination of ionization potential depression
  measurements.
\newblock {\em High Energy Density Physics}, 12:5--11, 2014.

\bibitem{vinko2014density}
SM~Vinko, O~Ciricosta, and JS~Wark.
\newblock Density functional theory calculations of continuum lowering in
  strongly coupled plasmas.
\newblock {\em Nature communications}, 5(1):3533, 2014.

\bibitem{Gu2008}
M~F Gu.
\newblock The flexible atomic code.
\newblock {\em Canadian Journal of Physics}, 86(5):675--689, 2008.

\bibitem{Gu2020}
M.~F. Gu and P.~Beiersdorfer.
\newblock Stark shift and width of x-ray lines from highly charged ions in
  dense plasmas.
\newblock {\em Phys. Rev. A}, 101:032501, Mar 2020.

\bibitem{rosmej2007}
{Rosmej, F. B.} and {Lee, R. W.}
\newblock Hollow ion emission driven by pulsed intense x-ray fields.
\newblock {\em EPL}, 77(2):24001, 2007.

\bibitem{perezcallejo-2023-dielectronic}
G.~Pérez-Callejo, T.~Gawne, T.~R. Preston, P.~Hollebon, O.~S. Humphries, H.~K.
  Chung, G.~L. Dakovski, J.~Krzywinski, M.~P. Minitti, T.~Burian,
  J.~Chalupský, V.~Hájková, L.~Juha, V.~Vozda, U.~Zastrau, S.~M. Vinko,
  S.~J. Rose, and J.~S. Wark.
\newblock Dielectronic satellite emission from a solid-density mg plasma:
  relationship to models of ionisation potential depression, 2023.

\bibitem{ZastrauHED}
Ulf Zastrau, Karen Appel, Carsten Baehtz, Oliver Baehr, Lewis Batchelor,
  Andreas Bergh{\"{a}}user, Mohammadreza Banjafar, Erik Brambrink, Valerio
  Cerantola, Thomas~E. Cowan, Horst Damker, Steffen Dietrich, Samuele
  Di~Dio~Cafiso, J{\"{o}}rn Dreyer, Hans-Olaf Engel, Thomas Feldmann, Stefan
  Findeisen, Manon Foese, Daniel Fulla-Marsa, Sebastian G{\"{o}}de, Mohammed
  Hassan, Jens Hauser, Thomas Herrmannsd{\"{o}}rfer, Hauke H{\"{o}}ppner,
  Johannes Kaa, Peter Kaever, Klaus Kn{\"{o}}fel, Zuzana Kon{\^{o}}pkov{\'{a}},
  Alejandro Laso~Garc{\'\i}a, Hanns-Peter Liermann, Jona Mainberger, Mikako
  Makita, Eike-Christian Martens, Emma~E. McBride, Dominik M{\"{o}}ller,
  Motoaki Nakatsutsumi, Alexander Pelka, Christian Plueckthun, Clemens
  Prescher, Thomas~R. Preston, Michael R{\"{o}}per, Andreas Schmidt, Wolfgang
  Seidel, Jan-Patrick Schwinkendorf, Markus~O. Schoelmerich, Ulrich Schramm,
  Andreas Schropp, Cornelius Strohm, Konstantin Sukharnikov, Peter Talkovski,
  Ian Thorpe, Monika Toncian, Toma Toncian, Lennart Wollenweber, Shingo
  Yamamoto, and Thomas Tschentscher.
\newblock {The High Energy Density Scientific Instrument at the European XFEL}.
\newblock {\em Journal of Synchrotron Radiation}, 28(5):1393--1416, Sep 2021.

\bibitem{Gawne2023}
Thomas Gawne, Thomas Campbell, Alessandro Forte, Patrick Hollebon, Gabriel
  Perez-Callejo, Oliver~S. Humphries, Oliver Karnbach, Muhammad~F. Kasim,
  Thomas~R. Preston, Hae~Ja Lee, Alan Miscampbell, Quincy~Y. van~den Berg, Bob
  Nagler, Shenyuan Ren, Ryan~B. Royle, Justin~S. Wark, and Sam~M. Vinko.
\newblock Investigating mechanisms of state localization in highly ionized
  dense plasmas.
\newblock {\em Physical Review E}, 108:035210, 9 2023.

\bibitem{Chung2017}
H.~K. Chung, B.~I. Cho, O.~Ciricosta, S.~M. Vinko, J.~S. Wark, and R.~W. Lee.
\newblock Atomic processes modeling of x-ray free electron laser produced
  plasmas using scfly code.
\newblock volume 1811. American Institute of Physics Inc., 3 2017.

\bibitem{Ren2023nat_comm}
Shenyuan Ren, Yuanfeng Shi, Quincy~Y. van~den Berg, Muhammad~F. Kasim,
  Hyun-Kyung Chung, Elisa~V. Fernandez-Tello, Pedro Velarde, Justin~S. Wark,
  and Sam~M. Vinko.
\newblock Non-thermal evolution of dense plasmas driven by intense x-ray
  fields.
\newblock {\em Communications Physics}, 6(1):99, May 2023.

\bibitem{Dornheim2022}
Tobias Dornheim, Maximilian B{\"o}hme, Dominik Kraus, Tilo D{\"o}ppner,
  Thomas~R. Preston, Zhandos~A. Moldabekov, and Jan Vorberger.
\newblock Accurate temperature diagnostics for matter under extreme conditions.
\newblock {\em Nature Communications}, 13(1):7911, Dec 2022.

\bibitem{xfel2806}
Katerina Falk, Šmíd Michal, and Baehtz Carsten.
\newblock Spectroscopic investigation of atomic physics in isochorically heated
  cu.
\newblock https://in.xfel.eu/metadata/doi/10.22003/XFEL.EU-DATA-002806-00,
  2022.

\bibitem{rodare_data}
Šmíd Michal.
\newblock Spectra from 2806 experiment at european xfel.
\newblock https://doi.org/10.14278/rodare.2789, 2024.

\bibitem{rodare_scripts}
Šmíd Michal~et al.
\newblock Analysis scripts for processing the 2806 experiment at european xfel.
\newblock https://doi.org/10.14278/rodare.2791, 2024.

\bibitem{Preston_2020}
T.R. Preston, S.~Göde, J.-P. Schwinkendorf, K.~Appel, E.~Brambrink,
  V.~Cerantola, H.~Höppner, M.~Makita, A.~Pelka, C.~Prescher, K.~Sukharnikov,
  A.~Schmidt, I.~Thorpe, T.~Toncian, A.~Amouretti, D.~Chekrygina, R.W. Falcone,
  K.~Falk, L.B. Fletcher, E.~Galtier, M.~Harmand, N.J. Hartley, S.P. Hau-Riege,
  P.~Heimann, L.G. Huang, O.S. Humphries, O.~Karnbach, D.~Kraus, H.J. Lee,
  B.~Nagler, S.~Ren, A.K. Schuster, M.~Smid, K.~Voigt, M.~Zhang, and
  U.~Zastrau.
\newblock Design and performance characterisation of the hapg von hámos
  spectrometer at the high energy density instrument of the european xfel.
\newblock {\em Journal of Instrumentation}, 15(11):P11033, nov 2020.

\bibitem{Pan}
X.~Pan, M.~Šmíd, R.~Štefaníková, F.~Donat, C.~Baehtz, T.~Burian,
  V.~Cerantola, L.~Gaus, O.~S. Humphries, V.~Hajkova, L.~Juha, M.~Krupka,
  M.~Kozlová, Z.~Konopkova, T.~R. Preston, L.~Wollenweber, U.~Zastrau, and
  K.~Falk.
\newblock {Imaging x-ray spectrometer at the high energy density instrument of
  the European x-ray free electron laser}.
\newblock {\em Review of Scientific Instruments}, 94(3):033501, 03 2023.

\bibitem{Chalupsky2015}
J.~Chalupsk\'y, P.~Boh\'a\ifmmode~\check{c}\else \v{c}\fi{}ek, T.~Burian,
  V.~H\'ajkov\'a, S.~P. Hau-Riege, P.~A. Heimann, L.~Juha, M.~Messerschmidt,
  S.~P. Moeller, B.~Nagler, M.~Rowen, W.~F. Schlotter, M.~L. Swiggers, J.~J.
  Turner, and J.~Krzywinski.
\newblock Imprinting a focused x-ray laser beam to measure its full spatial
  characteristics.
\newblock {\em Phys. Rev. Appl.}, 4:014004, Jul 2015.

\bibitem{Dornheim2023}
Tobias Dornheim, Maximilian~P. Böhme, David~A. Chapman, Dominik Kraus,
  Thomas~R. Preston, Zhandos~A. Moldabekov, Niclas Schlünzen, Attila Cangi,
  Tilo Döppner, and Jan Vorberger.
\newblock {Imaginary-time correlation function thermometry: A new,
  high-accuracy and model-free temperature analysis technique for x-ray Thomson
  scattering data}.
\newblock {\em Physics of Plasmas}, 30(4):042707, 04 2023.

\bibitem{Hell}
N.~{Hell}, G.~V. {Brown}, J.~{Wilms}, V.~{Grinberg}, J.~{Clementson},
  D.~{Liedahl}, F.~S. {Porter}, R.~L. {Kelley}, C.~A. {Kilbourne}, and
  P.~{Beiersdorfer}.
\newblock {Laboratory Measurements of the K-shell Transition Energies in L-
  shell Ions of SI and S}.
\newblock {\em \apj}, 830(1):26, October 2016.

\bibitem{Steinbrugge_2022}
René Steinbrügge, Steffen Kühn, Fabrizio Nicastro, Ming~Feng Gu, Moto
  Togawa, Moritz Hoesch, Jörn Seltmann, Ilya Sergeev, Florian Trinter, Sonja
  Bernitt, Chintan Shah, Maurice~A. Leutenegger, and José R.~Crespo
  López-Urrutia.
\newblock X-ray photoabsorption of density-sensitive metastable states in ne
  vii, fe xxii, and fe xxiii.
\newblock {\em The Astrophysical Journal}, 941(2):188, dec 2022.

\bibitem{Le2019}
Hai~P. Le, Mark Sherlock, and Howard~A. Scott.
\newblock Influence of atomic kinetics on inverse bremsstrahlung heating and
  nonlocal thermal transport.
\newblock {\em Phys. Rev. E}, 100:013202, Jul 2019.

\bibitem{Hau-Riege2013}
Stefan~P. Hau-Riege.
\newblock Nonequilibrium electron dynamics in materials driven by
  high-intensity x-ray pulses.
\newblock {\em Phys. Rev. E}, 87:053102, May 2013.

\end{thebibliography}


\appendix

\section{Methods}
\subsection{Experimental details}
\begin{figure}
    \centering
    \includegraphics[width=1\columnwidth]{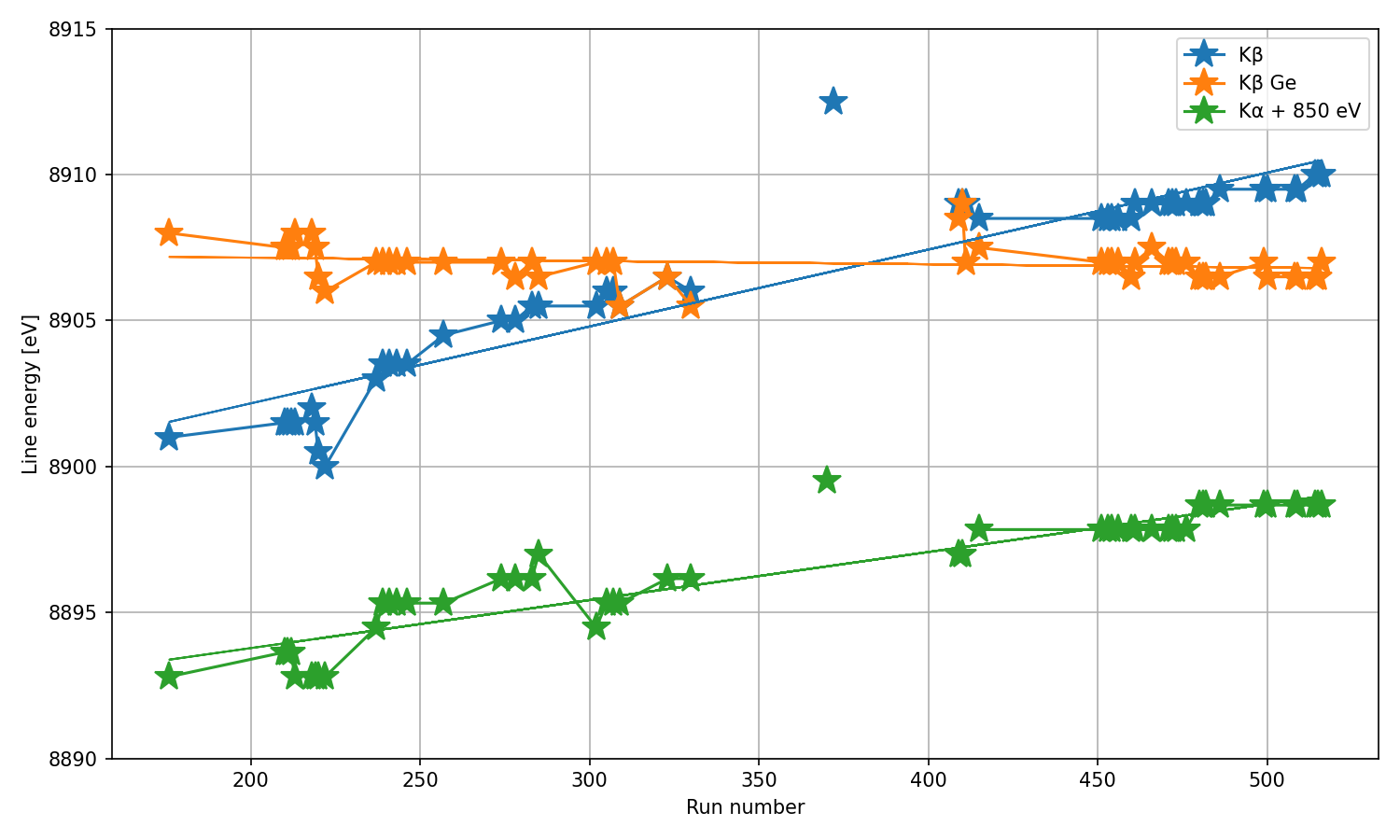}
    \caption{Calibration of spectrometers. Fitted position of \ka\ or \kb\ on the three used spectrometers as a function of run number, showing its shift during the progress of the experiment. The shown linear fit was used to correct for this shift. (Figure will be yet visually improved). }
    \label{calibration}
\end{figure}

The XFEL beam was operated in the SASE regime and was focused with a stack of 20 Beryllium lenses with radius of curvature 50~\um, providing a focal length of  $\approx$ 30 - 40 cm, based on the XFEL photon energy. 
The bandwidth was measured via scattering on cold sample and had a FWHM about 15 eV. The photon energy was scanned in the  range 8750  - 9900 eV with 25 eV steps. 
The x-ray emission was measured by three crystal spectrometers. Two of them employed the HAPG crystal and were aligned to the range 7900 - 8800 eV (measuring at scattering angle 35° - forward) and 8950 - 9750~eV (scattering angle 170° - backward), respectively \cite{Preston_2020}. The third one employed a Germanium crystal and observed the details of Cu \kb\ emission in the range 8950 - 9400 eV with higher resolution \cite{Pan}. The spectrometers were initially calibrated by measuring the emission of non-heated lines of Cu, Zn and Ni. However, during data analysis it was found the calibration of the HAPG spectrometer was shifting throughout the experiment, most likely due to unknown mechanical issue. This is shown in \figref{calibration}, where the results of the fits of Cu \ka\ and \kb\ are shown as a function of run number -- i.e. during the experimental progress, approximately 2 days. This shift was then fitted (straight lines) and all data were corrected by the found offset. The accuracy of spectral calibration is therefore better than 3 eV.

The energy of the XFEL beam was measured by fitting the XRTS peak visible in the spectrometers. The precision of the measurement is limited by the width of the SASE spectrum and spectrometer calibration, but can be estimated as better then 5 eV.

The target consists of simple foils held in a 6 mm $\times$ 30~mm window, allowing a continuous shooting with repetition frequency 10 Hz, when the speed of the target holder was adjusted to keep the spacing between shots 20 \um. The data in this paper are from shots where target was 3 \um\ thick Cu foil.

\subsection{Focusing}
\begin{figure}
    \centering
    \includegraphics[width=1\columnwidth]{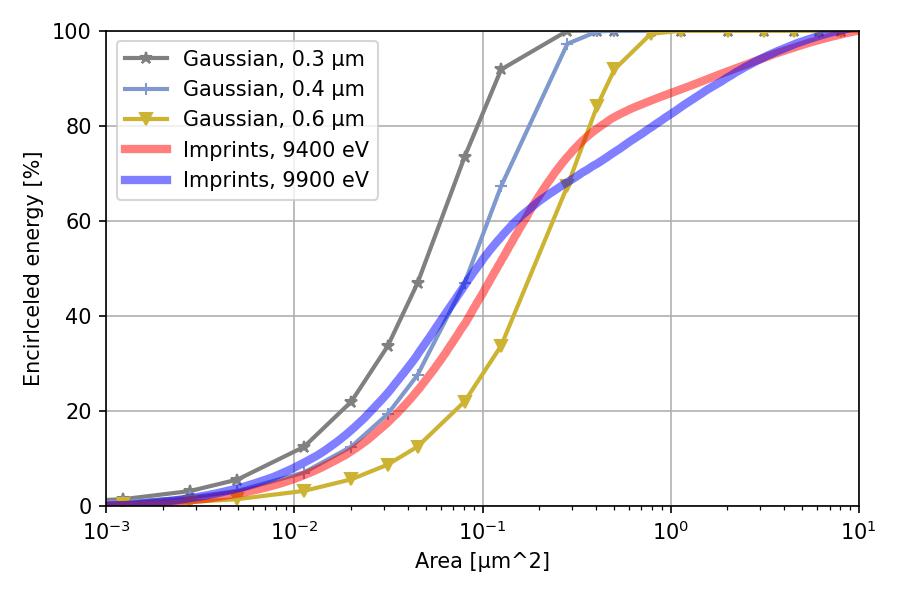}
    \caption{Focal spot characteristics (encircled energy) from the imprint measurement. Thin lines are calculations of 2D Gaussian spots for comparison.}
    \label{imprints}
\end{figure}

The focal length of the lenses is changing due to its chromacity by approximately 1 mm per 25 eV. Therefore, the focusing had to be optimized after each change of energy. This was done via continuous data acquisition while the XFEL was on and the lens position was changing. From such data, the position where strongest emission of ionized \ka\ lines was observed was identified as the focused one. The quality and characteristics of the focus was analyzed by the imprinting technique \cite{Chalupsky2015} at three photon energies (8900, 9400, and 9900 eV). The analysis of the imprints provided the encircled energy curves, \figref{imprints}. At all three cases it had shown comparable results, that the distribution of inner 50\% of energy was resembling 0.4~\um\ diameter Gaussian spot. Such size of the focus was mainly limited by the bandwidth of the beam. 

\subsection{Beam energy and intensity calibration}
\label{abel}
The beam energy on the target was measured by a diode detector coupled to a diamond screen, located between the last focusing element and the target. This detector was absolutely calibrated with the x-ray gas monitors. The last lens has an aperture of 300 \um, and was intentionally overfilled, i.e. the beam size on its entrance was kept in the range of approx. 350 - 500 \um. This overfilling converted the spatial jitter of the XFEL beam into an energy jitter, which means that in each data run, there was a large fluctuation of energy in individual shots, ranging between  30 to 280 \uJ. Each data run consists of typically 3000 shots, with nominally identical conditions. Due to the energy jitter, those shots have been grouped based on the measured energy providing energy resolved spectra. 

On those data, the so called \emph{focal spot inversion} was performed. This relies on the fact that experimental spectra for each beam energy, $S_E$, are integration of constituent spectra from various intensities, with ratios given by the focal spot distribution $f_I$, 
\[S_E(E) = \int S_I(I) f_I(I) dI, \]
where $I$ is the XFEL beam intensity, and $S_I$ is the emitted spectrum for given intensity. When considering a Gaussian focal spot profile, 
\begin{equation}
    I(r)=I_0\exp\left(-\frac{r^2}{2\sigma^2}\right),
\end{equation}
and considering the area enclosing each intensity contour $A=\pi r^2$, it is possible to construct the areal density
\begin{align}
    I(A) &= I_0e^{-\frac{A}{2\pi \sigma^2}}\\
    A(I) &= \left\{\begin{array}{lr}2\pi\sigma^2\ln\left(I_0/I\right) & 0<I<I_0 \\ 0 & \mathrm{otherwise}\end{array}\right.\\
    \frac{dA}{dI} &= \left\{\begin{array}{lr}-\frac{2\pi\sigma^2}{I} & 0<I<I_0 \\ 0 & \mathrm{otherwise}\end{array}\right.\\
\end{align}
from which it is possible to construct the inverse mapping
\begin{align}
    S_E(E)&=\int_0^{I_0}S_I(I)\left|\frac{dA}{dI}\right|dI\\
    &=\int_0^{I_0}\frac{2\pi\sigma^2S_I(I)}{I}dI\\
    I_0 &= \frac{E}f{2\pi\sigma^2}\\
    \frac{dS_E(E)}{dE} &= \frac{S_I(I)}{I}.
\end{align}

The discretized form of the differential was used to obtain the spectra presented in this paper \[S_I (I)/I = \frac{d S_E}{d E},\] where we retain the normalization by $I$ to yield the emitted spectrum for a given incident power at the specified intensity, so that line emission intensity is comparable.

While, as shown in \figref{imprints}, the focal spot is not purely Gaussian, the most intense parts of the distribution -- which is responsible for the heated emission -- is well described by a Gaussian profile. This could therefore be expected to induce errors on reconstructed emission from ground state emission lines, \ka\ and \kb, which have contributions from the large wings of the focal spot profile -- however does not affect the conclusions of this work on the plasma screening within solid density HED plasmas.

\subsection{Spectra fitting}
\label{fitting}
The measured spectra were fitted with a set of gaussian curves. The fit minimizes the least squares of differences between the experimental spectrum and fitted curve, but due to a high non-linearity of the problem, a custom algorithm was applied.

\begin{figure} [hbt]
\begin{center}
\includegraphics[width=\columnwidth]{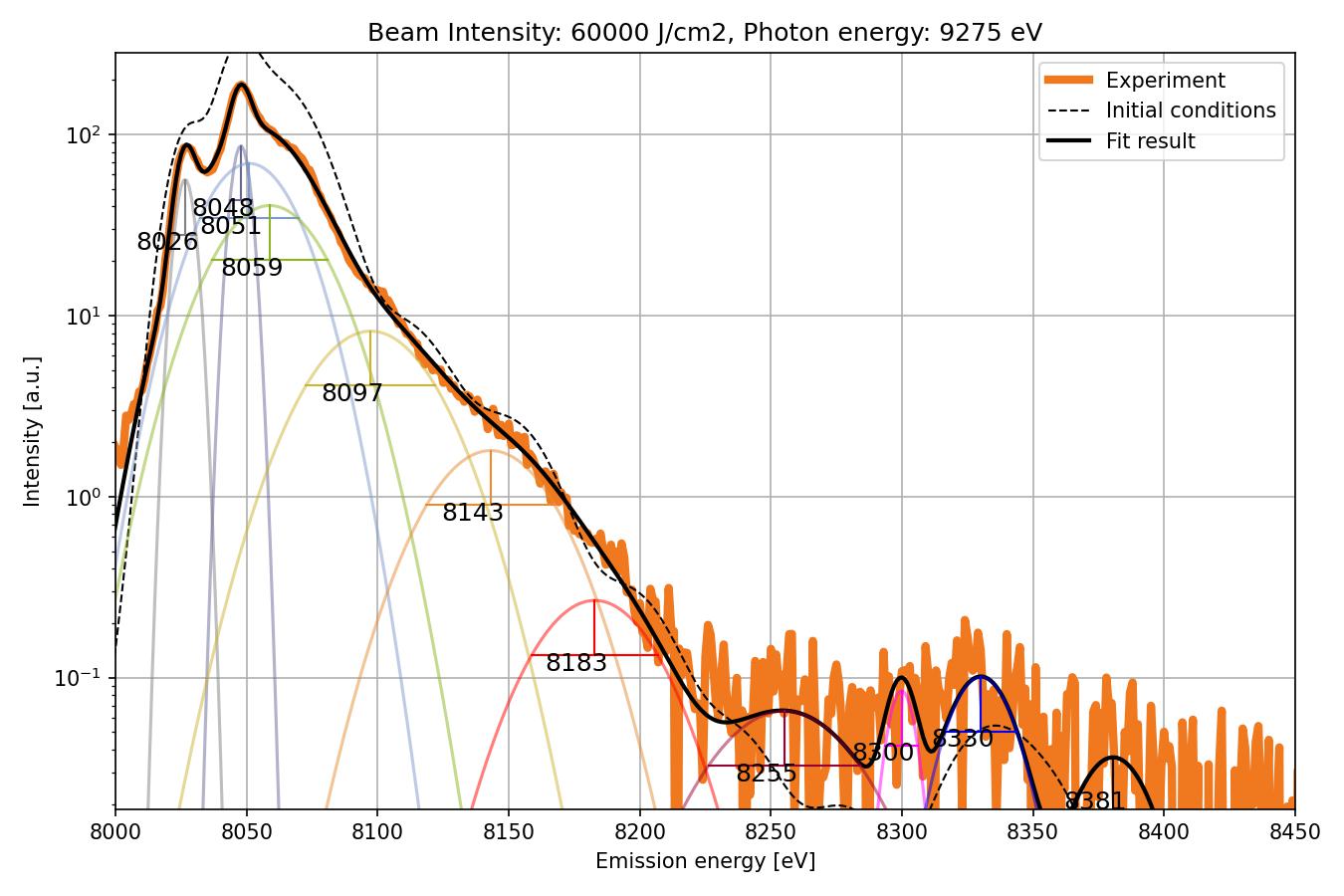}

\includegraphics[width=\columnwidth]{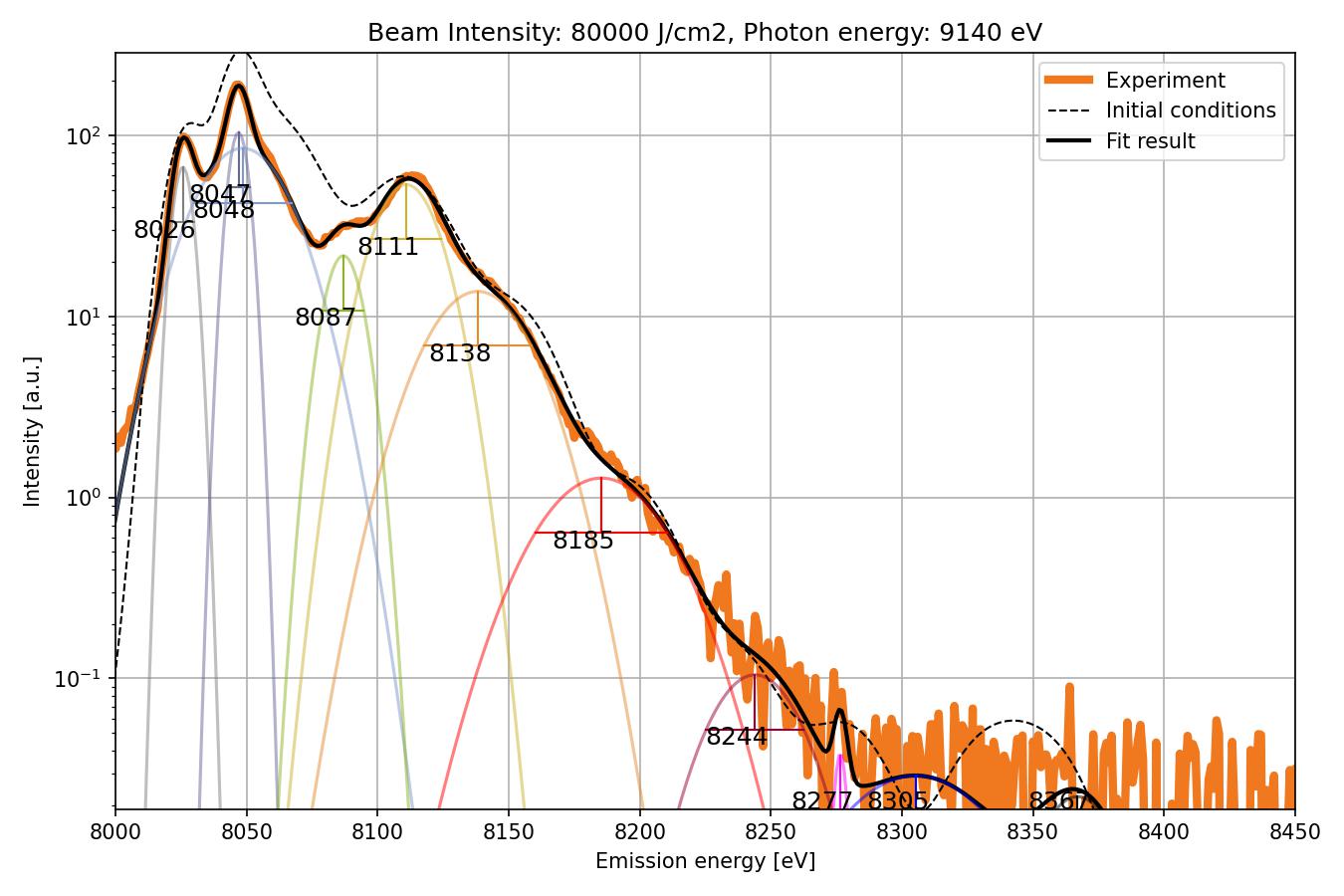}

\includegraphics[width=\columnwidth]{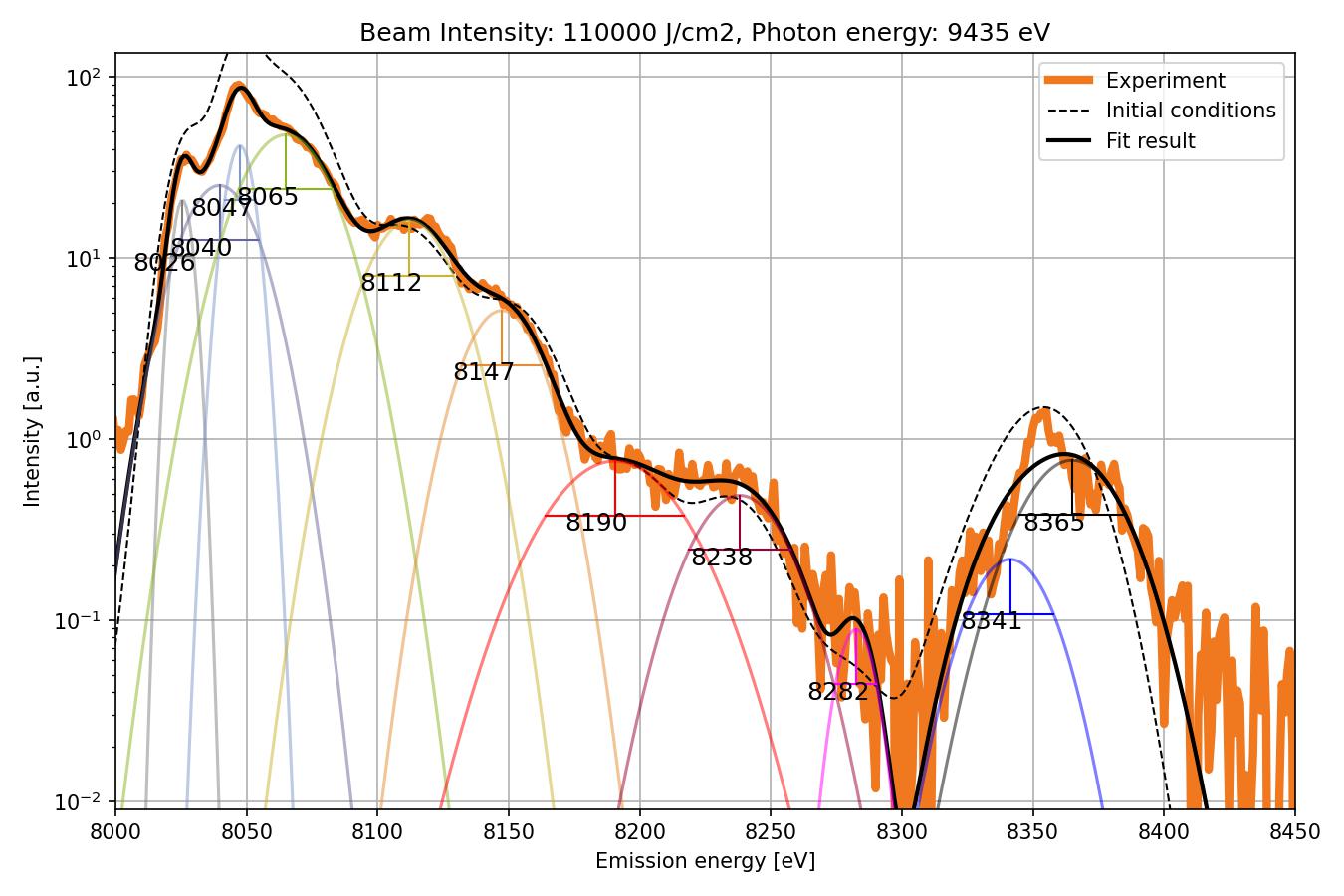}

\caption{\label{fitKA} Selected experimental spectra in the \ka\ range and their fits for few selected irradiation conditions, see respective titles. The colored parabolas depicts the gaussian components of the fits with label stating fitted peak energy.}
\end{center}
\end{figure}

In order to catch the properties of very low intensity lines comparably well to the strong ones, the fit is perfomed in the $log$ space. The algorithm is iterative. For the \ka\ spectra, it is fitting sum of 11 gaussian lines. In order to gain convergence, a proper initial conditions have to be set. The initial line positions are $8027, 8047, 8055, 8065, 8110, 8150, 8195, 8235, 8278, 8332$ and $8355$ eV, the initial widths are 15 eV for the cold \ka , and 35 eV for the satellites, the amplitudes are set to correspond to experimental values at the positions. The fit is then minimizing the difference in the range 8010 --  8400 eV. The components with initial conditions between 8065 eV and 8278 eV corresponds to the L8 -- L3 transitions. Components with higher initial energy fill the \kah\ range, which is then fitted separately. The spectra, initial conditions, and fit results including separate components are shown in \figref{fitKA}.

\begin{figure} [hbt]
\begin{center}
\includegraphics[width=\columnwidth]{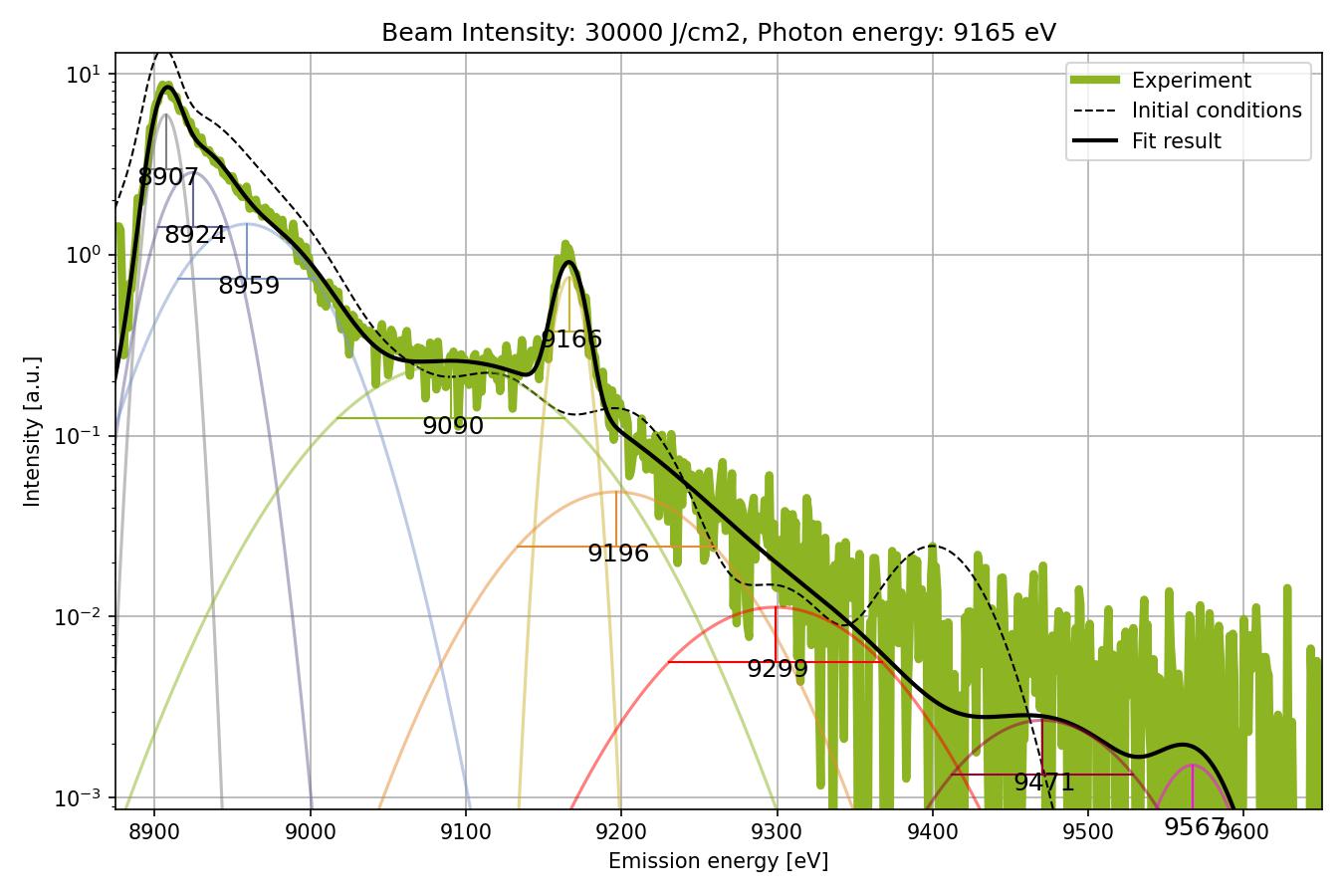}

\includegraphics[width=\columnwidth]{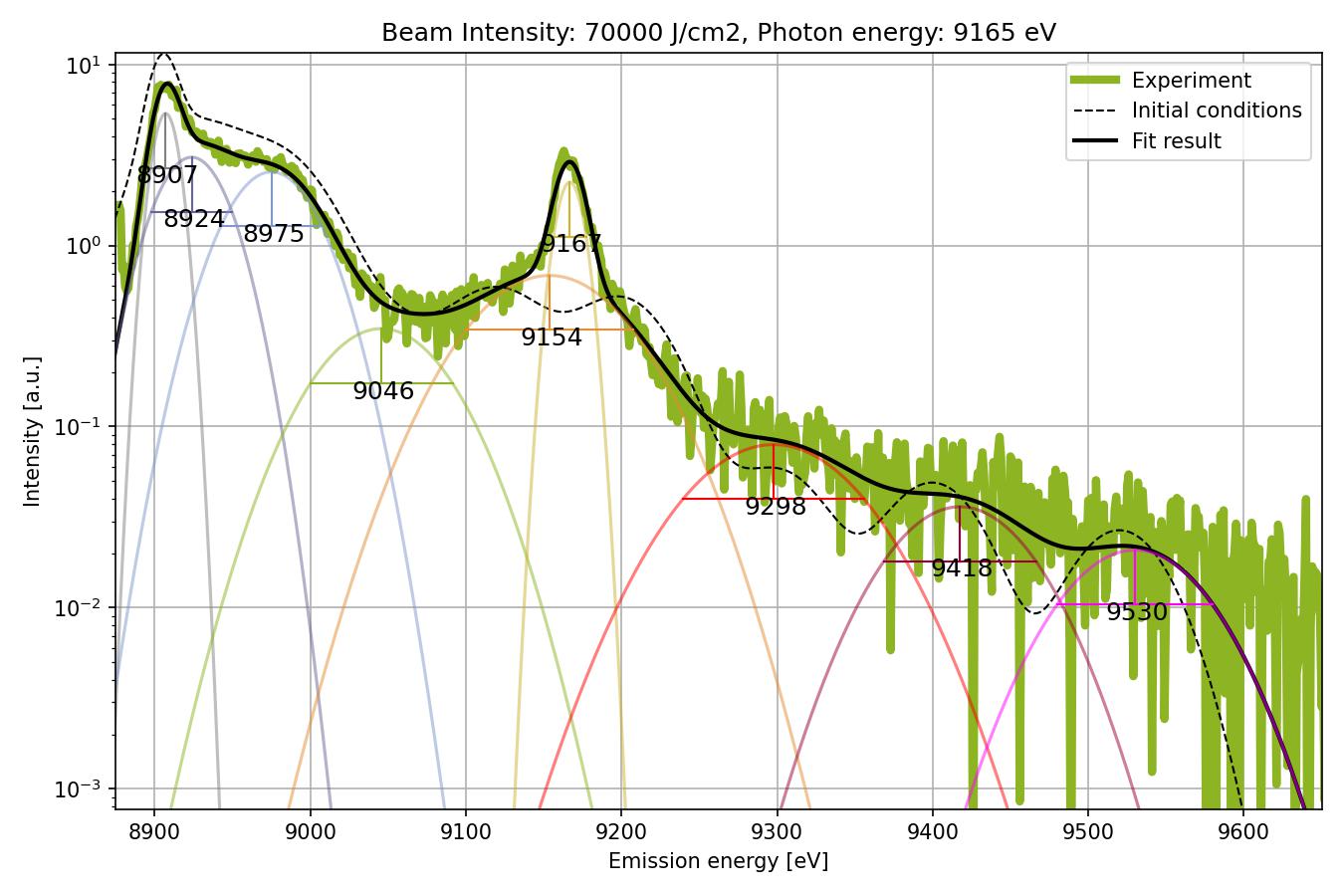}

\includegraphics[width=\columnwidth]{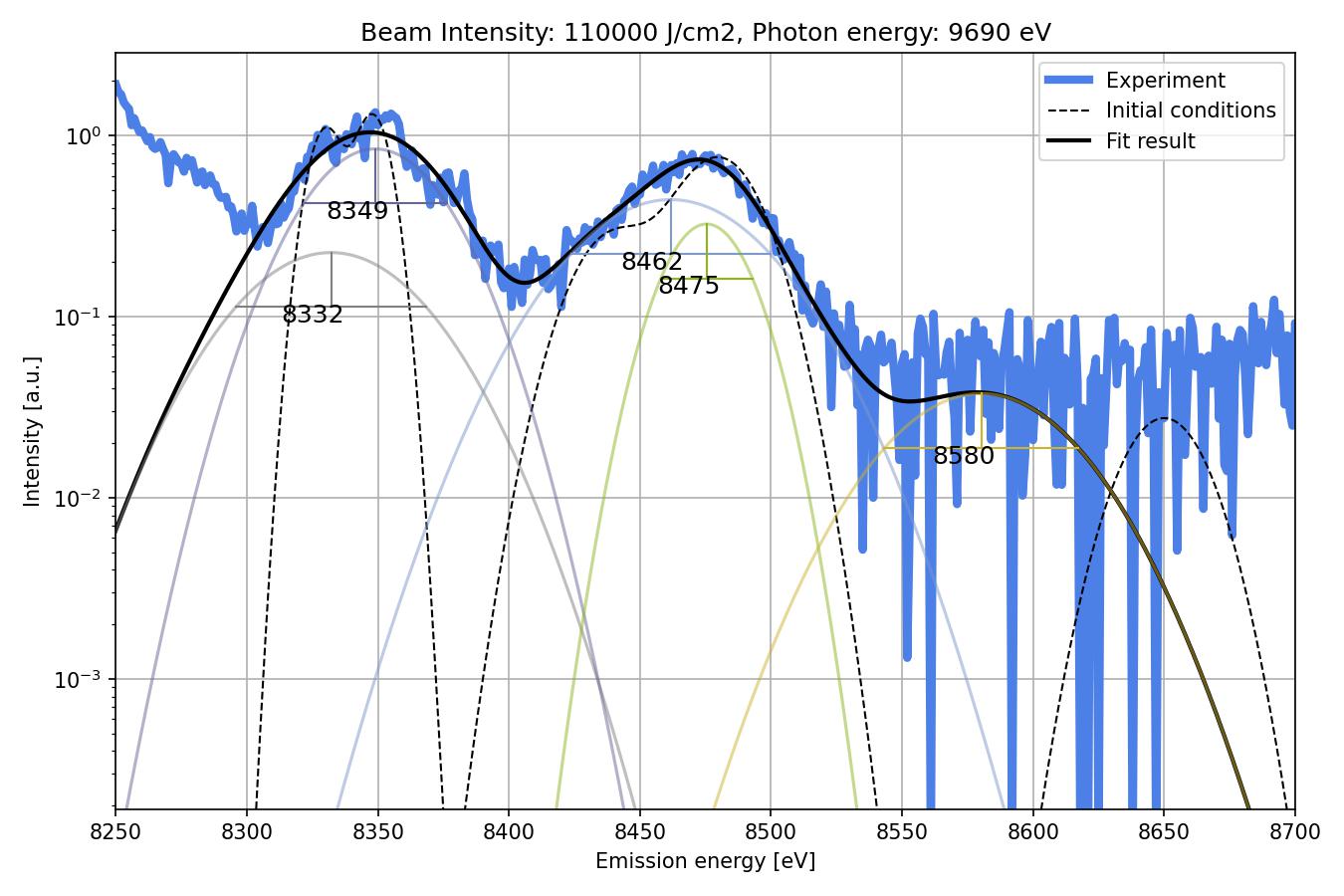}

\caption{\label{fitKB} Experimental spectra in the \kb\ and \kah\ range and their fits for few selected irradiation conditions, see respective titles. The colored parabolas depicts the gaussian components of the fits with label stating fitted peak energy.}
\end{center}
\end{figure}

The same algorithm is then used to fit the \kb\ and \kah\ range. For \kb, the range 8800 -- 9600 eV is fitted, and the components are starting at  $8905, 8920, 8980, 9050, 9120, 9200, 9300, 9400$ and $9520$ eV. For \kah , the fitted range is 8300 -- 8600 eV and the initial conditions are $8330, 8348, 8440, 8480$  and $8650$ eV. The spectra and results are shown in \figref{fitKB}

\subsection{Shift fitting}
Measuring the shift of each line as a difference of theoretical and experimental position can be misleading due to often complicated line shapes. As already, for example, the essential \ka\ is formed by a doublet with $\approx 20$ eV spacing. In the theoretical data, the weaker component is seen and affects the mean (central) position of the peak, while in complex experimental spectra might be neglected or confused with emission of neighboring charge states.

\begin{figure} [hbt]
\begin{center}

\includegraphics[width=\columnwidth]{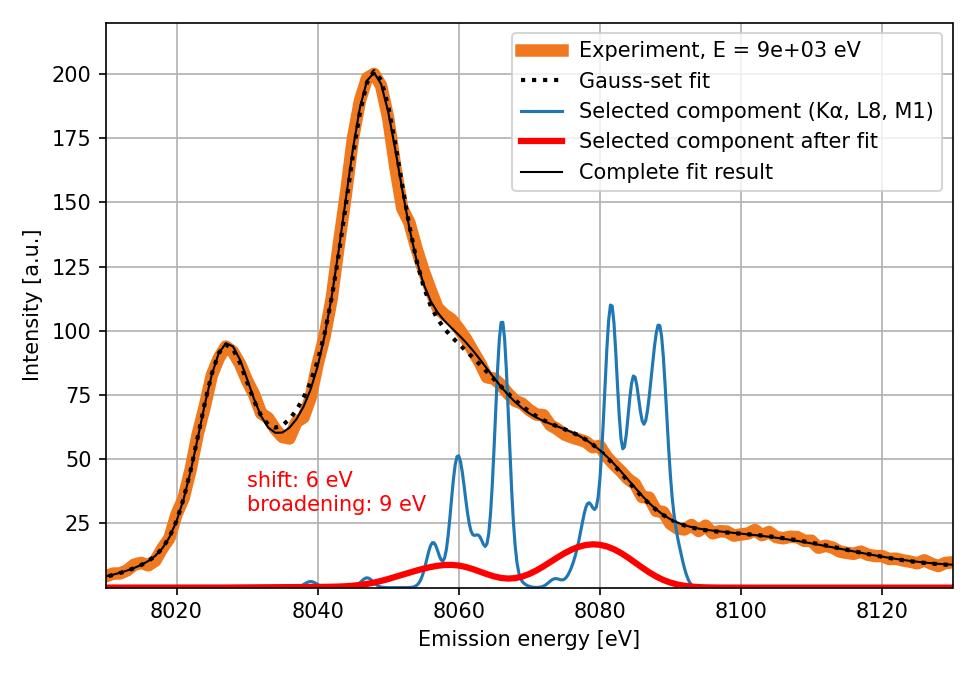}

\caption{\label{finefit} Illustration of fitting of shift of single component.}
\end{center}
\end{figure}

To overcome this, we have developed following approach to fit the shift of each component by using the complex line shape as predicted by the simulation. This process is performed for each identified transition separately, and is illustrated in \figref{finefit} on example of \ka\ in Z=19 with 8 electrons in L shell. First, the fit of experimental spectra by a set of gaussian components is taken, as decribed in previous section (dotted line). The gaussian component corresponding to the line of interested is then removed and replaced by the spectrum of given line, as calculated by the FAC code (blue line). The shift, amplitude, and broadening of this component is fitted via least square method to get best agreement (black line) to the experimental curve. 

The chosen example shows that the stronger, higher-energy component of the transition (8080 - 8090 eV) is well fit to the shoulder on experimental data at 8080 eV, while the other, low energy part of spectra, is still seen in the data at 8060 eV. This fit result in shift of 6 eV of this component, which is directly shown in \figref{screening}(a).

\subsection{XRTS}
\label{xrts}
\begin{figure} [hbt]
\begin{center}
\includegraphics[width=0.8\columnwidth]{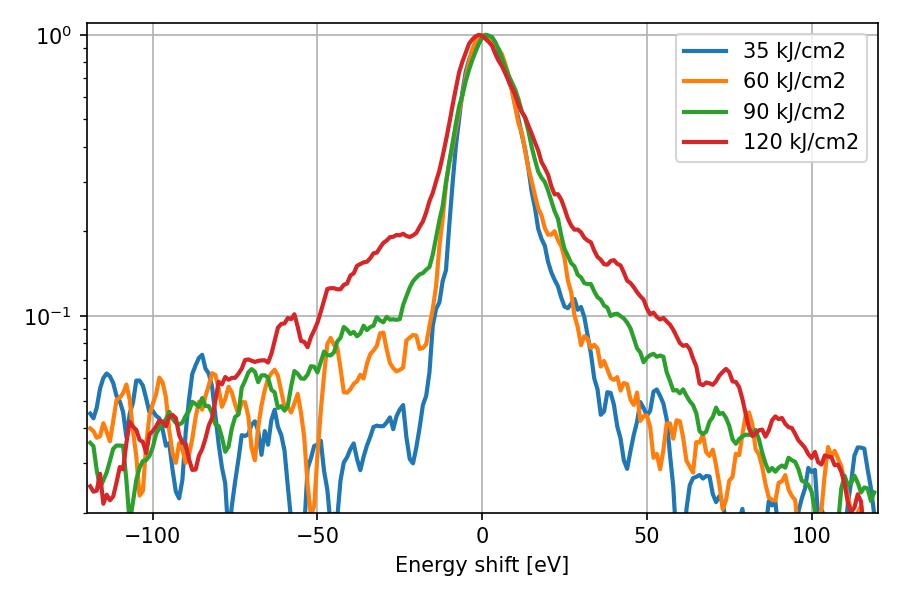}
\caption{\label{figxrts} Spectra used for the XRTS analysis}
\end{center}
\end{figure}

The XRTS spectrum measured at scattering angle 170° (see \figref{figxrts}), was analysed by the imaginary time thermometry method~\cite{Dornheim2022, Dornheim2023}.
The XRTS spectrum and the source and instrument function are subjected to a two-sided Laplace transformation. The resulting quotient of the two quantities is symmetric in imaginary time $\tau$-space around the inverse temperature $k_B T/2$ due to detailed balance. This allows to extract the temperature model-free. 

\begin{figure}[hbtp]
    \centering
    \includegraphics[width=\linewidth]{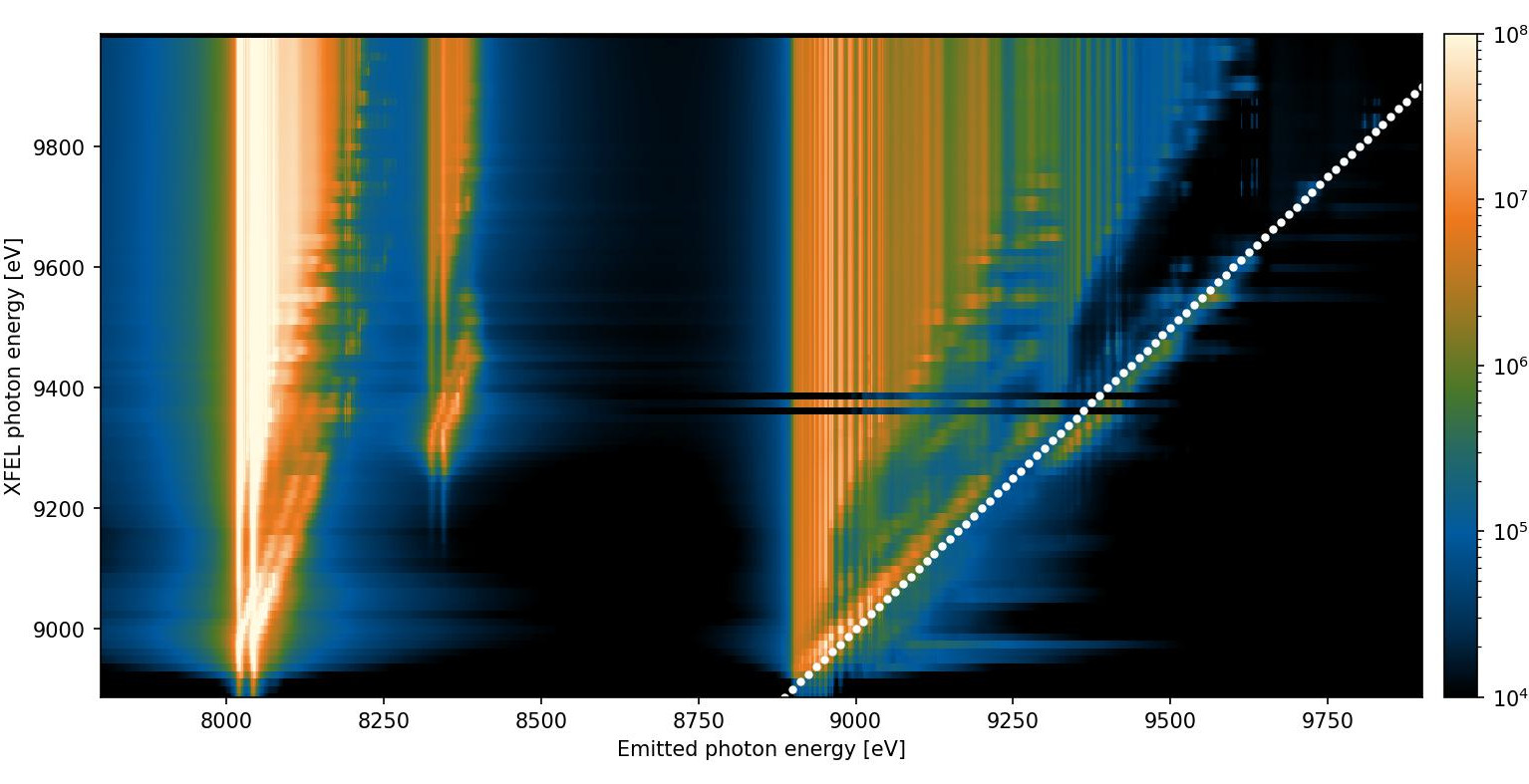    }
    \caption{Spectra simulated by the SCFLY code for irradiation 125 \kjcm.}
    \label{scfly}
\end{figure}

\subsection{Atomic simulations}
\label{app.fac}
The FAC code was used to calculate the energies of x-ray lines and edges. In this case, we have very higher control of the atomic model, and using the new functionality of FAC \cite{Gu2020}, we can vary the plasma screening model and see its effect on those. Results of those calculations are seen in Fig.2 and described in the main text. The input files for the model are generated by a script which looks for levels needed to produce the desired transitions. There are typically tens of configurations per charge state, the whole model then has about  226 000 levels.

The accuracy of line position calculation in isolated ion is critical for the analysis in this paper. The accuracy of FAC calculations on Si and S was assesed in \cite{Hell}, showing a very good agreement on a 2-3 eV level. In Cu, the EBIT measurements were compared to the second order many-body perturbation theory (MBPT mode) FAC calculations yielding typically 0.5 eV accuracy \cite{Steinbrugge_2022}. That is however a more involved mode, which may not be possible for very complex ions with many open M-shell electrons as are in this work. Calculations in this paper are therefore done in the configuration interaction (CI) mode only. To benchmark those, a simulations in identical settings but on Fe were performed and compared to the experimental data presented in \cite{Steinbrugge_2022}. The results are shown in Tab. \ref{accuracy}, showing a $<2$\ eV accuracy. This is much smaller then scale of line shift discussed in this paper (tens eV).

\begin{table}[hbt]
    \centering  
    \begin{tabular}{lcccc}
    \hline\hline 
    Element & Line  & $E_\mathrm{exp}$ [eV] & $E_\mathrm{FAC}$ [eV] & $\Delta$ [eV] \\
    \hline
        Fe XXII & C1 & 6544.2 & 6542.8 & 1.4\\
		Fe XXII & C2 & 6556.9 & 6554.5 & 2.4\\
    \hline
    \end{tabular}
    \caption{Comparison of line positions from our atomic calculations  ($E_\mathrm{FAC}$) to experimental values presented in Table 1 of \cite{Steinbrugge_2022} ($E_\mathrm{exp}$) and their difference.}
    \label{accuracy}
\end{table}

Figure \ref{FAC2} shows calculations for further lines with no plasma screening. Figure \ref{FAC-SP} shows calculations for \ka\ line with plasma screening following the Stewart-Pyatt model for various assumptions of electron temperature.



\subsection{Collisional radiative simulations}
\label{app-scfly}
The SCFLY code was used to model the interaction in a  0D, time-dependent collisional-radiative (CR) simulations with zero initial temperature, and with heating by XFEL beam with super Gaussian temporal profile. We have performed a set of simulations with various beam and photon energies, to simulate the whole set of experimental data, see \figref{scfly}. We can observe the simulation matches the experiment qualitatively very well, but such approach is not suitable for quantitative analysis as the atomic model lacks the necessary details. 

\subsection{Cretin non-thermal CR simulations}
\label{cretin}
The Cretin simulations model the evolution of the electron distribution and do not assume a thermal distribution. It evolves the electron energy distribution self-consistently with the atomic populations. The electron distribution evolution is governed by a kinetic Boltzmann-Fokker-Planck equation which includes all elastic (electron -- electron collision) and inelastic collisional and radiative processes \cite{Le2019}. The atomic model used in the Cretin simulations has the same set of energy levels and transitions as SCFly. Since the model does not assume thermal distribution, the temperature shown in \figref{temperatures}(a) is defined as 2/3 of the kinetic energy of the electrons. Deviation from a thermal distribution can be calculated using the non-equilibrium factor defined by Hau-Riege \cite{Hau-Riege2013}. First, the energy is decomposed into a low-energy and a high energy component. The non-equilibrium factor is defined as the ratio of the kinetic energy of the high energy component relative to the total kinetic energy, i.e., 0 means thermalized distribution and 1 means complete non-equilibrium distribution. This factor is decreasing below 0.1 towards the end of XFEL pulse, showing the plasma is sufficiently collisional to be modelled as thermalized.

\section{Extended data}

\subsection{Resonances}
The identified resonances  are written as the pairs of driving (absorption) and emission energy in Tab.\ref{resonances}. The uncertainty of emission energy is written in table, the absorption uncertainty is given by the 25 eV bandwidth of the XFEL. Second section of the table assign identified processes and charge state to the pair. Last column classifies certainty of this assignment: 3 - confident identification, 2 - less confident identification, 1 - estimate, 0 - no suitable identification was found. Last section shows the result of shift fitting using the FAC line shape, as discused above. The fitted shift and broadeding is stated.

\setlength{\tabcolsep}{3pt}
{\makeatletter
\renewcommand\table@hook{\small}
\makeatother

\begin{longtable*}[p]{c|lll|llllc|ll}%
  \# &
 \multicolumn{3}{c|}{Energy [eV]} &
 \multicolumn{5}{c|}{Identification} &
  \multicolumn{2}{c}{Fit [eV]}\\
 &  Absorption & Emission & Unc. & 
 Z & L & Process & States & Cert.&
  Shift & FWHM
  \\ \hline
\begin{filecontents*}{figures/16_table.csv}
4,8960, 8057, 1.5 ,15,8, B - A,K2 L8 M4 - K1 L8 M5 - K2 L7 M5,1, , 
5,9165, 8059, 2.5 ,14,8, G - A,K2 L8 M5 - K1 L8 M5 N1 - K2 L7 M5 N1,1, , 
8,9220, 8068, 2.5 ,16,8, G - A,K2 L8 M3 - K1 L8 M3 N1 - K2 L7 M3 N1,2, , 
9,9010, 8069, 1.5 ,17,8, B - A,K2 L8 M2 - K1 L8 M3 - K2 L7 M3,2,1,14
11,9040, 8073, 1.5 ,18,8, B - A,K2 L8 M1 - K1 L8 M2 - K2 L7 M2,2, , 
12,9275, 8073, 2.5 ,17,8, G - A,K2 L8 M2 - K1 L8 M2 N1 - K2 L7 M2 N1,3,3,12
15,9320, 8080, 1.5 ,18,8, G - A,K2 L8 M1 - K1 L8 M1 N1 - K2 L7 M1 N1,2,6,9
19,9090, 8928, 5.0 ,11,8, G - B,K2 L8 M8 - K1 L8 M8 N1 - K2 L8 M7 N1,2,39,45
20,9115, 8945, 5.0 ,12,8, G - B,K2 L8 M7 - K1 L8 M7 N1 - K2 L8 M6 N1,1,45,49
22,9165, 8972, 5.0 ,14,8, G - B,K2 L8 M5 - K1 L8 M5 N1 - K2 L8 M4 N1,1, , 
25,9220, 9003, 7.5 ,16,8, G - B,K2 L8 M3 - K1 L8 M3 N1 - K2 L8 M2 N1,2, , 
27,9275, 9025, 12.5 ,17,8, G - B,K2 L8 M2 - K1 L8 M2 N1 - K2 L8 M1 N1,3, , 
29,8960, 9146, 20.0 ,13,8, B - G,K2 L8 M5 N1 - K1 L8 M6 N1 - K2 L8 M6,1, , 
30,9010, 9220, 10.0 ,16,8, B - G,K2 L8 M2 N1 - K1 L8 M3 N1 - K2 L8 M3,2,88,86
31,9040, 9255, 15.0 ,17,8, B - G,K2 L8 M1 N1 - K1 L8 M2 N1 - K2 L8 M2,3, , 
33,9320, 8328, 1.5 ,13,8, Bh - Ah,K1 L8 M7 - K0 L8 M8 - K1 L7 M8,1, , 
34,9370, 8340, 1.5 ,16,8, Bh - Ah,K1 L8 M4 - K0 L8 M5 - K1 L7 M5,1,11,64
35,9435, 8354, 1.0 ,19,8, Bh - Ah,K1 L8 M1 - K0 L8 M2 - K1 L7 M2,1,5,36
40,9090, 8103, 2.5 ,17,7, B - A,K2 L7 M3 - K1 L7 M4 - K2 L6 M4,2, , 
41,9345, 8104, 1.5 ,17,7, G - A,K2 L7 M3 - K1 L7 M3 N1 - K2 L6 M3 N1,2,4,29
42,9115, 8109, 2.5 ,18,7, B - A,K2 L7 M2 - K1 L7 M3 - K2 L6 M3,3,1,32
43,9385, 8110, 1.5 ,18,7, G - A,K2 L7 M2 - K1 L7 M2 N1 - K2 L6 M2 N1,2, , 
45,9140, 8113, 1.5 ,19,7, B - A,K2 L7 M1 - K1 L7 M2 - K2 L6 M2,3, , 
46,9165, 8118, 1.5 ,20,7, B - A,K2 L7 M0 - K1 L7 M1 - K2 L6 M1,3, , 
47,9535, 8154, 2.5 ,20,6, Bh - A,K1 L7 M1 - K0 L7 M2 - K1 L6 M2 - K2 L5 M2,3, , 
48,9560, 8156, 2.5 ,21,6, Bh - A,K1 L7 M0 - K0 L7 M1 - K1 L6 M1 - K2 L5 M1,1, , 
49,9535, 8395, 3.0 ,20,7, Bh - Ah,K1 L7 M1 - K0 L7 M2 - K1 L6 M2,2, , 
52,9360, 9113, 5.0 ,17,7, G - B,K2 L7 M3 - K1 L7 M3 N1 - K2 L7 M2 N1,3,30,62
54,9385, 9125, 7.5 ,18,7, G - B,K2 L7 M2 - K1 L7 M2 N1 - K2 L7 M1 N1,2,37,61
56,9140, 9420, 15.0 ,19,7, B - G,K2 L7 M0 N1 - K1 L7 M1 N1 - K2 L7 M1,3, , 
66,9195, 8142, 2.5 ,18,6, B - A,K2 L6 M3 - K1 L6 M4 - K2 L5 M4,3, , 
67,9220, 8151, 2.5 ,19,6, B - A,K2 L6 M2 - K1 L6 M3 - K2 L5 M3,3,1,21
68,9515, 8150, 1.5 ,19,6, G - A,K2 L6 M2 - K1 L6 M2 N1 - K2 L5 M2 N1,2,6,24
70,9245, 8154, 2.5 ,20,6, B - A,K2 L6 M1 - K1 L6 M2 - K2 L5 M2,3,4,29
71,9275, 8159, 2.5 ,21,6, B - A,K2 L6 M0 - K1 L6 M1 - K2 L5 M1,3, , 
72,9635, 8192, 1.5 ,21,5, Bh - A,K1 L6 M1 - K0 L6 M2 - K1 L5 M2 - K2 L4 M2,2, , 
73,9610, 8434, 1.5 ,20,6, Bh - Ah,K1 L6 M2 - K0 L6 M3 - K1 L5 M3,1, , 
74,9460, 9179, 12.5 ,16,6, G - B,K2 L6 M5 - K1 L6 M5 N1 - K2 L6 M4 N1,1, , 
75,9535, 9229, 7.5 ,20,6, G - B,K2 L6 M1 - K1 L6 M1 N1 - K2 L6 M0 N1,2, , 
76,9220, 9511, 17.5 ,19,6, B - G,K2 L6 M1 N1 - K1 L6 M2 N1 - K2 L6 M2,2, , 
77,9245, 9537, 15.0 ,20,6, B - G,K2 L6 M0 N1 - K1 L6 M1 N1 - K2 L6 M1,1, , 
83,9275, 8185, 3.5 ,19,5, B - A,K2 L5 M3 - K1 L5 M4 - K2 L4 M4,1, , 
87,9690, 8195, 1.5 ,22,5, Bh - A,K1 L5 M1 - K0 L5 M2 - K1 L5 M1 - K2 L4 M1,1,7,20
88,9345, 8195, 1.5 ,21,5, B - A,K2 L5 M1 - K1 L5 M2 - K2 L4 M2,2, , 
93,9370, 8198, 2.5 ,22,5, B - A,K2 L5 M0 - K1 L5 M1 - K2 L4 M1,2, , 
95,9690, 8474, 4.5 ,22,5, Bh - Ah,K1 L5 M1 - K0 L5 M2 - K1 L4 M2,1, , 
96,9610, 9295, 7.5 ,19,5, G - B,K2 L5 M3 - K1 L5 M3 N1 - K2 L5 M2 N1,2, , 
97,9635, 9319, 7.5 ,20,5, G - B,K2 L5 M2 - K1 L5 M2 N1 - K2 L5 M1 N1,2,31,74
98,9360, 9680, 20.0 ,21,5, B - G,K2 L5 M0 N1 - K1 L5 M1 N1 - K2 L5 M1,1, , 
99,9320, 9620, 15.0 ,20,5, B - G,K2 L5 M1 N1 - K1 L5 M2 N1 - K2 L5 M2,2, , 
103,9410, 8229, 2.5 ,22,4, B - A,K2 L4 M1 - K1 L4 M2 - K2 L3 M2,2,8,22
104,9460, 8232, 2.5 ,23,4, B - A,K2 L4 M0 - K1 L4 M1 - K2 L3 M1,2,9,25
107,9385, 9720, 17.5 ,21,4, B - G,K2 L4 M1 N1 - K1 L4 M2 N1 - K2 L4 M2,2, , 
108,9815, 8237, 1.5 ,24,4, Bh - A,K1 L4 M0 - K0 L4 M1 - K1 L4 M0 - K2 L3 M0,3,12,24
109,9765, 8490, 5.0 ,23,4, Bh - Ah,K1 L4 M1 - K0 L4 M2 - K1 L3 M2,1, , 
110,9815, 8518, 2.5 ,24,4, Bh - Ah,K1 L4 M0 - K0 L4 M1 - K1 L3 M1,3, , 
111,9815, 9423, 15.0 ,22,4, G - B,K2 L4 M1 - K1 L4 M1 N1 - K2 L4 M0 N1,2,44,107
115,9515, 8276, 2.0 ,23,3, B - A,K2 L3 M1 - K1 L3 M2 - K2 L2 M2,3, , 
116,9535, 8280, 1.5 ,24,3, B - A,K2 L3 M0 - K1 L3 M1 - K2 L2 M1,3, , 
117,9915, 8278, 2.0 ,25,3, Bh - A,K1 L3 M0 - K0 L3 M1 - K1 L3 M0 - K2 L2 M0,2, , 
118,9915, 8565, 2.5 ,25,3, Bh - Ah,K1 L3 M0 - K0 L3 M1 - K1 L2 M1,2, , 
122,9965, 8567, 3.5 ,25,2, Bh - Ah,K1 L2 M1 - K0 L2 M2 - K1 L1 M2,1, , 
123,9690, 8351, 2.5 ,26,1, B - A,K2 L1 M0 - K1 L1 M1 - K2 L0 M1,2, , 
\end{filecontents*}
\csvreader[head to column names]{figures/16_table.csv}{}
    {\csvcoli & \csvcolii & \csvcoliii & \csvcoliv 
    & \csvcolv  & \csvcolvi & \footnotesize{\csvcolvii} & \csvcolviii  & \csvcolix   & \csvcolx & \csvcolxi \\}
\\
    \caption{Measured resonances and their identification for  $I = 110 $\ \kjcm . }
        
    \label{resonances}
\end{longtable*}}

\newcommand\x{7}

\begin{figure*}[p]
    \centering
    \includegraphics[width=\x cm]{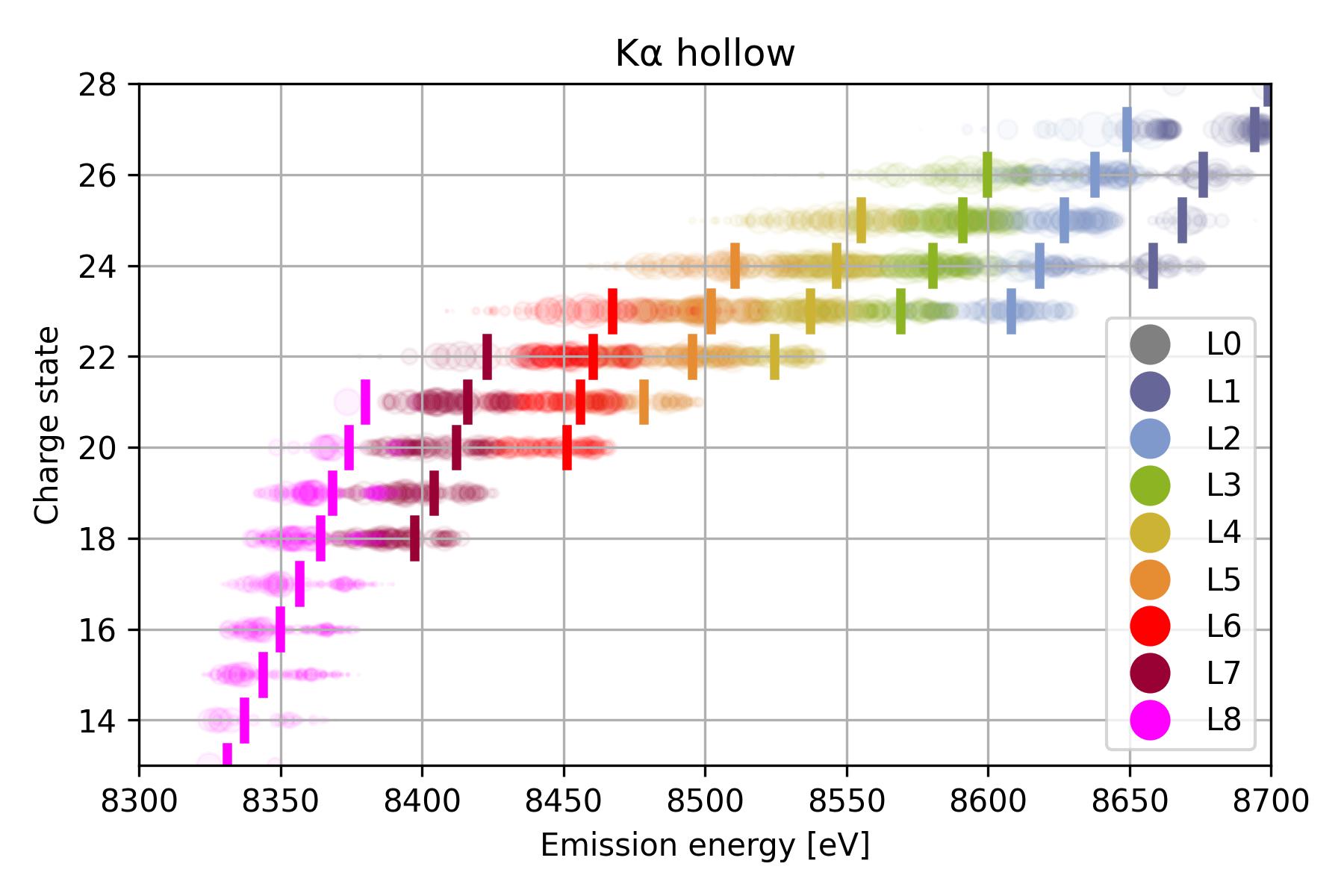}     
    \includegraphics[width=\x cm]{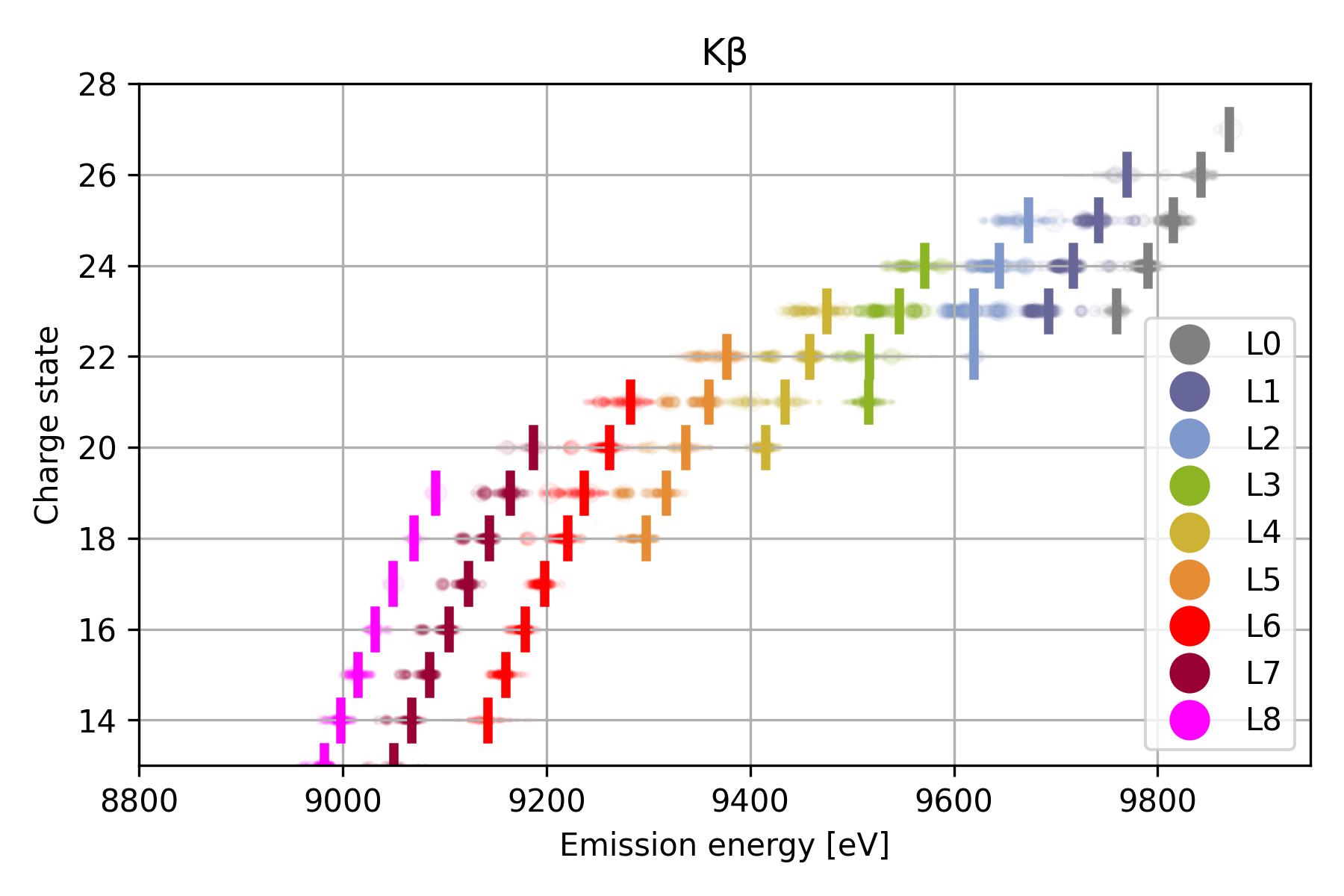}    
    \includegraphics[width=\x cm]{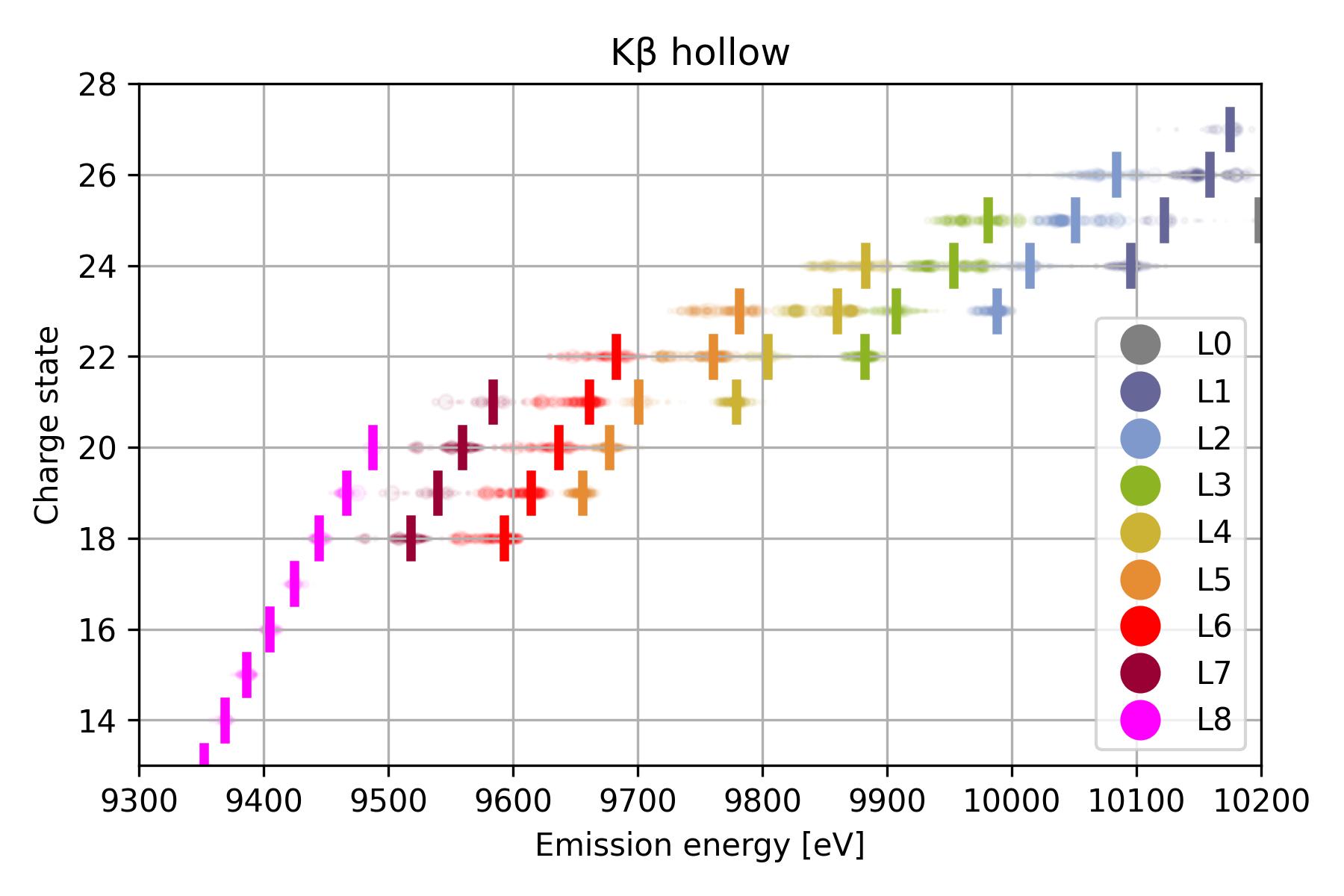}    
    \includegraphics[width=\x cm]{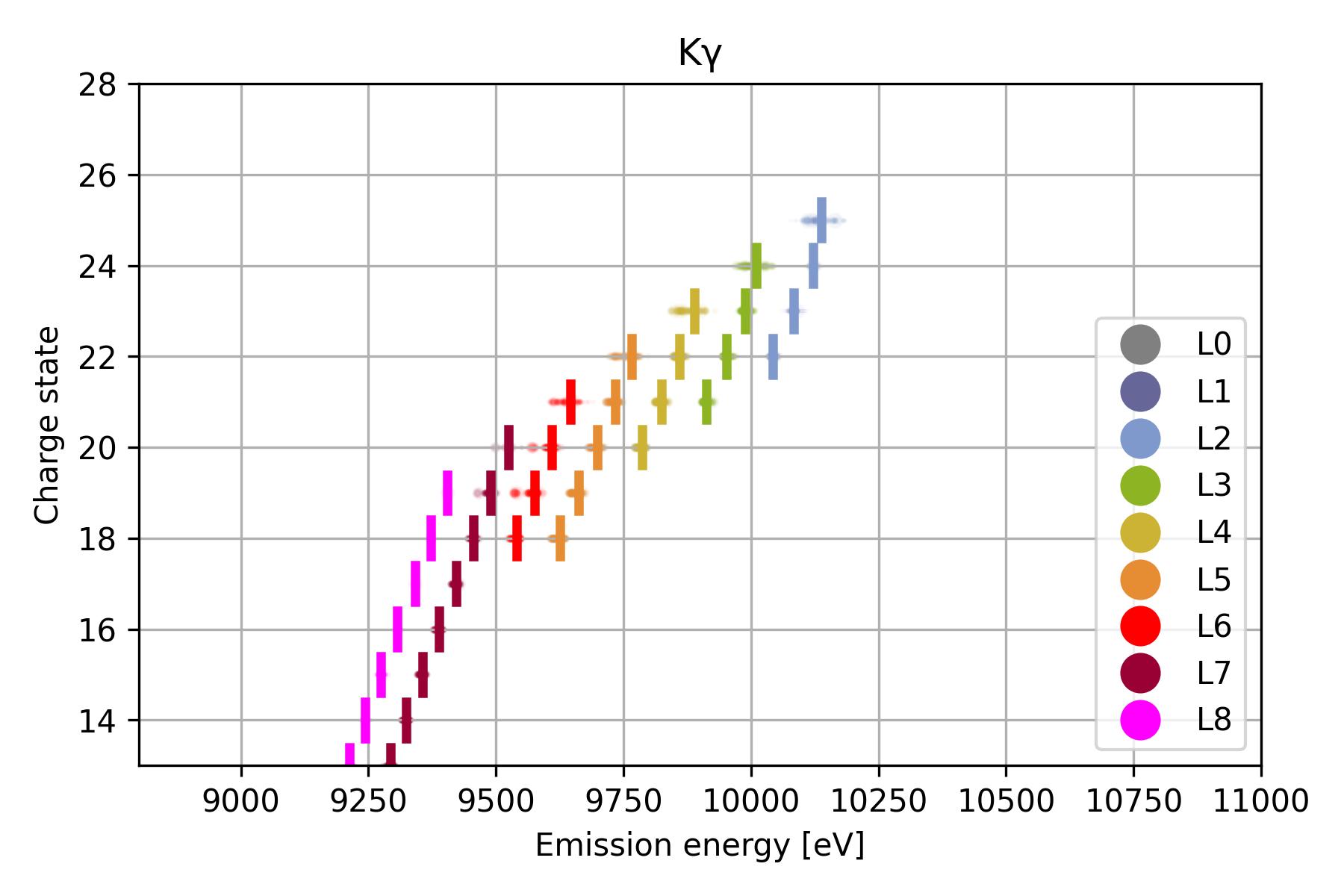}
    
    \caption{Line positions calculated by FAC for various lines, see title of respective subfigures.}
    \label{FAC2}
\end{figure*}

\begin{figure*}[p]
    \includegraphics[width=\x cm]{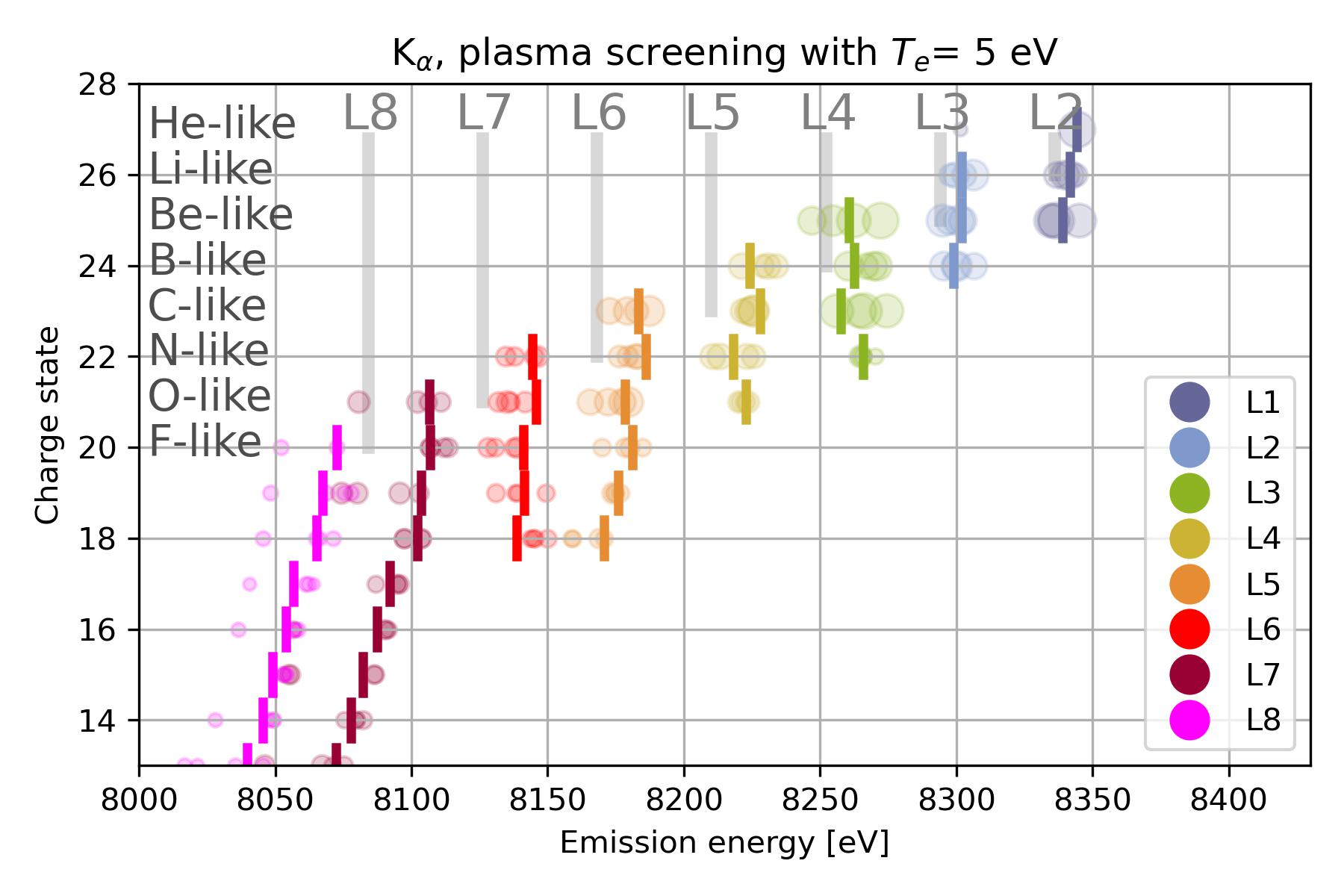}
    \includegraphics[width=\x cm]{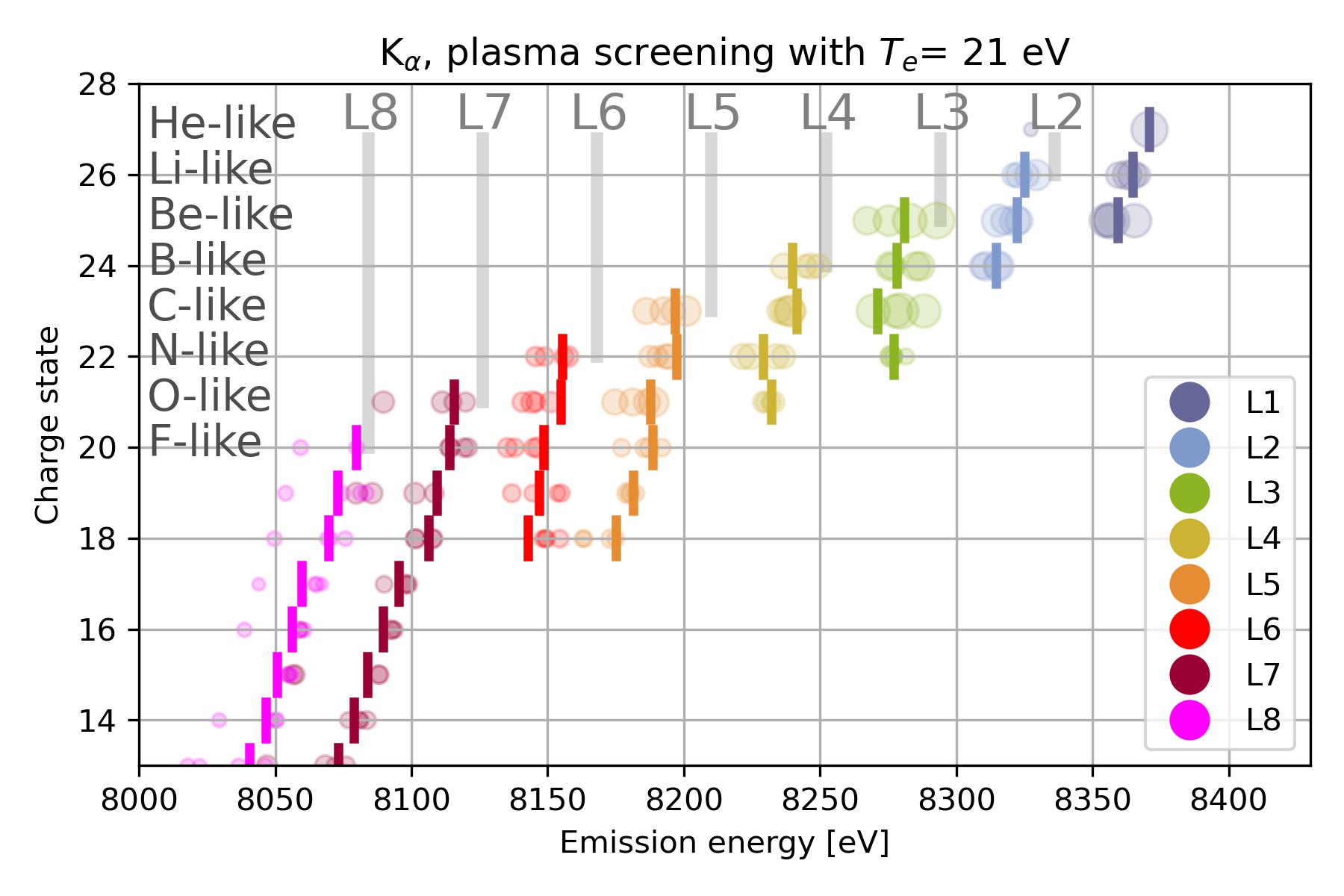}
    \includegraphics[width=\x cm]{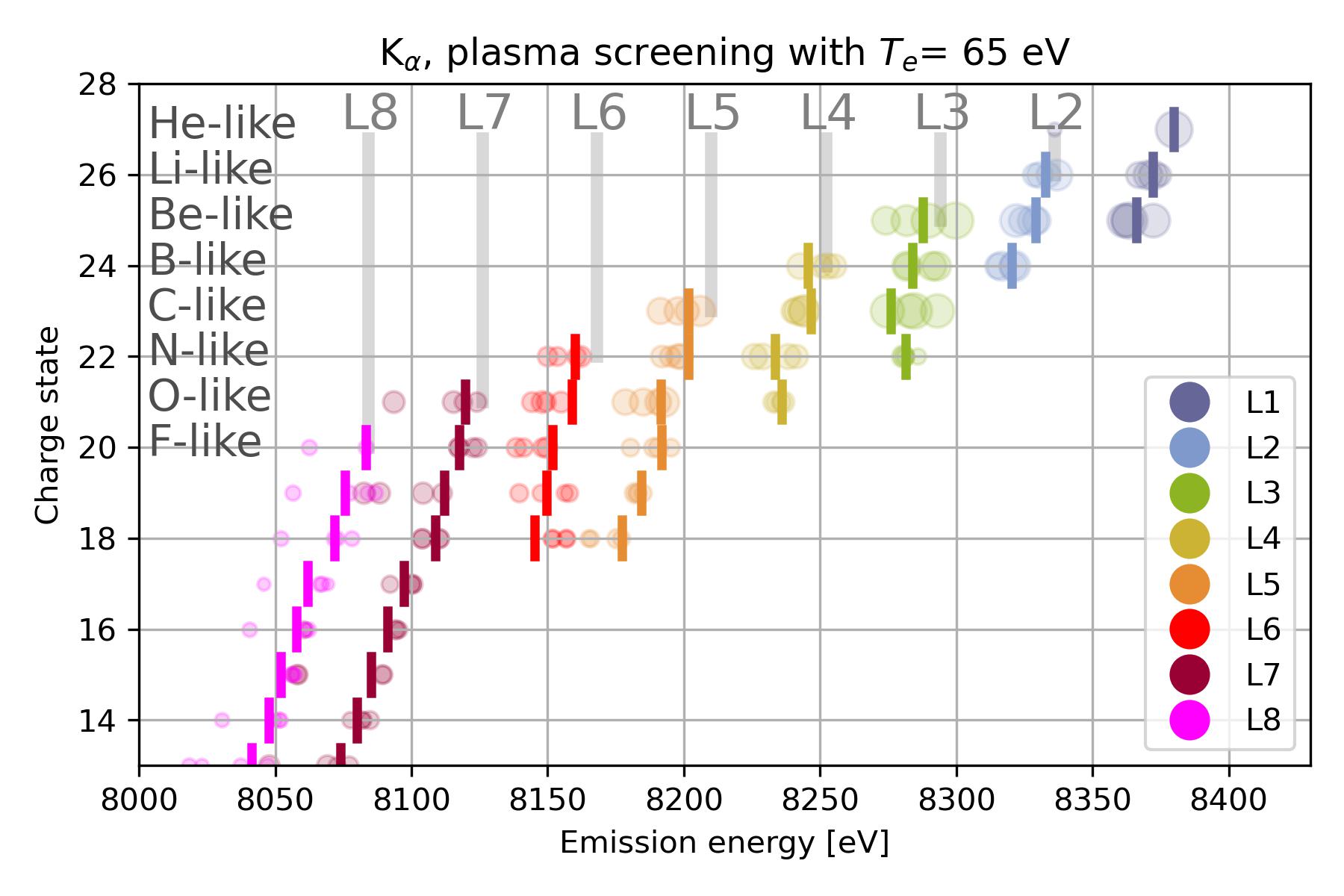}
    \includegraphics[width=\x cm]{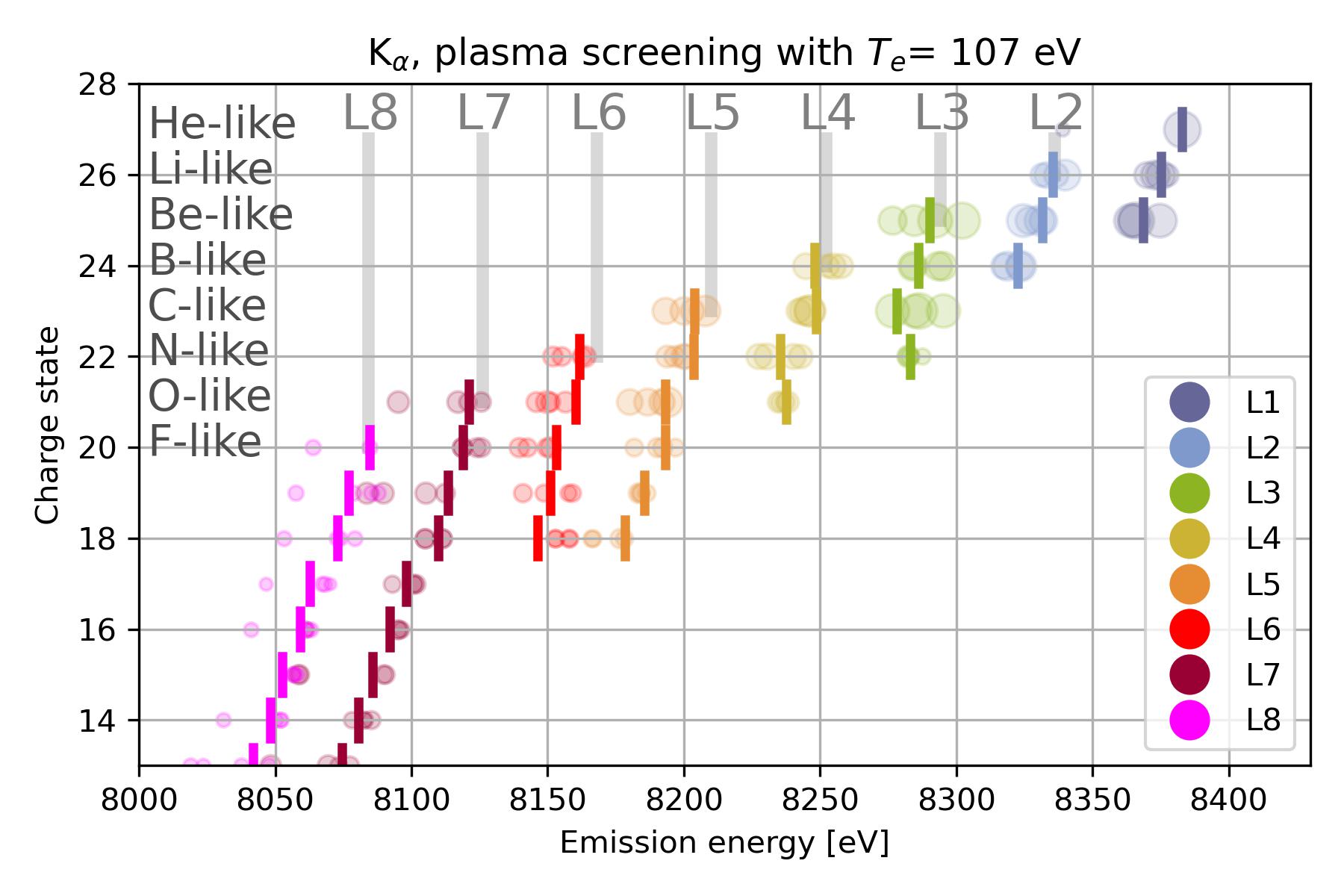}

\caption{Energies of \ka\ calculated by FAC with Stewart Pyatt model of plasma screening with various temperature assumptions, see respective titles.}
    \label{FAC-SP}
\end{figure*}

\end{document}